\documentclass{nature}
\usepackage{mathtools, cuted}
\usepackage[x11names]{xcolor}
\newcommand\abs[1]{\left|#1\right|}
\usepackage{comment}
\usepackage{lipsum}
\usepackage{adjustbox}
\usepackage{graphicx}
\usepackage{comment}
\usepackage{amsmath}
\usepackage{lipsum}
\usepackage{float}
\usepackage{nicefrac}
\usepackage{soul}
\usepackage{cancel}
\usepackage[normalem]{ulem}
\usepackage{multirow,tabularx}
\usepackage{dblfloatfix}
\usepackage{nidanfloat}
\usepackage{dcolumn}
\usepackage{bm}
\usepackage{hyperref}
\usepackage{upgreek}
\usepackage{amsmath}
\usepackage{lineno}

\usepackage{multicol}

\newcommand{\beginsupplement}{%
        \setcounter{table}{0}
        \renewcommand{\thetable}{S\arabic{table}}%
        \setcounter{figure}{0}
        \renewcommand{\thefigure}{S\arabic{figure}}%
        \setcounter{equation}{0}
        \renewcommand{\theequation}{S.\arabic{equation}}
        \setcounter{section}{0}
        \renewcommand{\thesection}{SM\arabic{section}}  
     }

\title{Large-scale Ising Emulation with Four-Body Interaction and All-to-All Connection}

\author{Santosh Kumar$^{1,2}$, He Zhang$^{1,2}$ \& Yu-Ping Huang$^{1,2}$}%

\begin{document}
\maketitle

\begin{affiliations}
 \item Department of Physics, Stevens Institute of Technology, Hoboken, NJ, 07030, USA
 \item Center for Quantum Science and Engineering, Stevens Institute of Technology, Hoboken, NJ, 07030, USA
\end{affiliations}

 \date{12/26/2019}
  

\begin{abstract}
We propose and demonstrate a nonlinear optics approach to emulate Ising machines containing up to a million spins and with tailored two and four-body interactions with all-to-all connections. It uses a spatial light modulator to encode and control the spins in the form of the binary phase values of wavelets in coherent laser beams, and emulates the high-order interaction with frequency conversion in a nonlinear crystal at the Fourier plane. By adaptive feedback control, the system can be evolved into effective spin configurations that well approximate the ground states of Ising Hamiltonians with all-to-all connected many-body interactions. Our technique could serve as a new tool to probe complex, many-body physics and give rise to exciting applications in big data optimization, computing, and analytics.  
\end{abstract}


\maketitle


\section{\label{sec:level1} Introduction}
A wide range of modern applications across biology \cite{liu_dna_2000}, medicine \cite{granan_ising_2016}, finance \cite{Mantegna:1999}, and social networks \cite{stauffer_social_2008} benefit from efficient processing and optimization of big data with complex structures and correlations. However, many such tasks are non-deterministic polynomial time hard (NP-hard), which could take existing supercomputers years to solve \cite{Garey:1990}. In this challenge, intense research efforts are underway to pave alternative approaches for computing and information processing. Among them, Ising machines have been shown to offer viable solutions for important NP-hard problems such as MAX-CUT \cite{Haribara2016824}, protein folding \cite{aksel_analysis_2009}, and traveling salesman \cite{applegate_traveling_2007}, among others \cite{lucas_ising_2014,xia_electronic_2018,Goto2019,chou_analog_2019,Yoshioka2019,Dlaska2019,Hamerly2019}. To this end, a variety of Ising machines have been demonstrated in effective spin systems of trapped atoms \cite{labuhn_tunable_2016,adams_rydberg_2019}, polariton condensates \cite{ohadi_spin_2017}, superconducting circuits \cite{heras_digital_2014}, coupled oscillators \cite{mahboob_electromechanical_2016,Wang_2019,bello_persistent_2019}, nanophotonic waveguides \cite{harris_quantum_2017,pitsios_photonic_2017,roques-carmes_integrated_2019}, randomly coupled lasers \cite{nixon_observing_2013,pal_observing_2017,babaeian_single_2019}, and time-multiplexed optical parametric oscillators \cite{inagaki_coherent_2016,Takesue2019}. 

For the tasks of finding the ground states of many-body Hamiltonians, photonic systems enjoy the distinct advantages of high connectivity and speed \cite{kirkpatrick_optimization_1983,santoro_theory_2002,puri_quantum_2017,xue_experimental_2017,Takata2014,Takata2016}. For example, a fast coherent Ising machine can be realized in a looped optical parametric oscillator with temporally multiplexed pulses \cite{Inagaki2016415,Haribara2016251}, albeit with limited spin numbers \cite{Inagaki2016415} or relying on photodetection and electronic feedback to emulate the spin-spin interaction \cite{Clements2017,bohm_poor_2019}. In contrast, a linear-optical Ising machine based on spatial light modulation was shown to subtend about 80,000 spins by coding them as the binary phases of pixels on a spatial light modulator (SLM) \cite{Pierangeli2019}. The far-field optical power of a modulated beam gives the expected energy of spin-spin interaction. The relatively simple setup yet high connectivity and scalability make this approach attractive to Ising machines with fully connected two-body interaction.

Yet, there are physical systems and numeric models whose dynamics cannot be fully captured by two-body interactions, and proper descriptions of multi-body interaction are required
\cite{Wu_PRB,Lieb72,PRB_Pretti,mizel_three-_2004, gurian_observation_2012}. This poses a significant computational challenge, whose complexity and volume exceeds by far that of Ising problems with only two-body interaction, even for a moderate number of spins \cite{georgescu_quantum_2014,Dai2017}. While a small class of many-body interaction can be decomposed onto a series of two-body interactions via some recursive or algebraic means \cite{biamonte_nonperturbative_2008,leib_transmon_2016,bernien_probing_2017}, they often subject to strict constraints \cite{bacon_universal_2000,divincenzo_universal_2000} or require tedious error corrections \cite{lechner_quantum_2015,lidar_quantum_2013}. For simulating complex systems and processing data with high-order correlation, suitable Ising machines remain desirable that support simultaneously high  connectivity, multi-body interaction, and a large number of spins.  


In this paper, we propose and experimentally demonstrate such an Ising machine hosting adjustable two-body interaction, four-body interaction, and all-to-all connections over a large number of spins. It realizes the spins as the binary phases of wavelets in a coherent laser beam, and implements effective multi-body interaction through nonlinear frequency conversion. Using SLM's (or equivalently, digital micromirror devices), one million spins are easily accessible. The fully connected two-body interaction is emulated with the optical power of the modulated light in the Fourier plane. The four-body interaction, also fully connected, is realized effectively by passing the modulated light through a lithium-niobate crystal in the Fourier plane for second harmonic (SH) generation. By simultaneously measuring the optical powers of the modulated light and its SH coupling into a fiber, complex Hamiltonians with all-to-all connected two-body and four-body interactions can be emulated over nearly one million spins. Through feedback control, the system can be evolved into the vicinity of the ground state of its Hamiltonian, exhibiting ferromagnetic, paramagnetic, and other novel nonlinear susceptibility phase transitions.

The present Ising emulator could pave a pathway to otherwise inaccessible territories of big data analytics and quantum simulation \cite{mizel_three-_2004,gurian_observation_2012}. The high-order, many-body interaction can also serve as powerful activation functions for all optical machine learning  \cite{zuo_all-optical_2019,miscuglio_all-optical_2018}. For example, an immediate application is to use this machine as the q-state Potts model with two-body and four-body interactions on a square lattice \cite{schreiber_ferromagnetic_2018}. Finally, while the current setup uses SH generation, even richer physics and controllability can be achieved by using other nonlinear optical processing like sum-frequency generation \cite{Santosh19}, and four-wave mixing \cite{four-wave_mixing}, where other types of spin interaction and connection can be engineered. 

\section{\label{sec:level2}Theoretical Analysis}
\begin{figure*}[htbp]
    \centering
    \includegraphics[width=\linewidth]{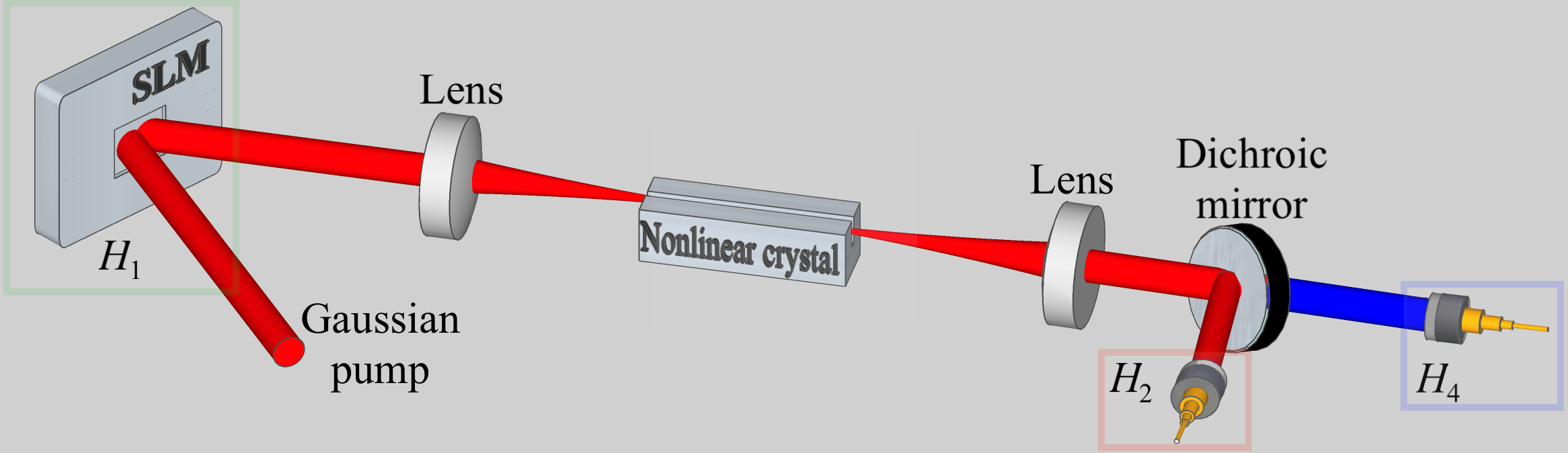}
    \caption{Illustration of an Ising machine with adjustable chemical energy, all-to-all two-body and four-body interactions. }
    \label{Sketch_2}
\end{figure*}

The basic idea of the present Ising machine is illustrated in Fig.~\ref{Sketch_2}, which emulates chemical potential, two-body interaction, and four-body interaction over a large number of spins. Each spin is encoded as the binary phase of a pixel on a SLM. The total chemical potential energy is represented by the weighted sum of all spins. To realize the interactions, a coherent Gaussian pump beam is reflected off the SLM, focused using a Fourier lens to a nonlinear crystal for SH generation. The resulting beams at the original fundamental wavelength and the new SH wavelength are then separated at a dichroic mirror and captured by optical fibers. The pump power in the fiber is then measured to emulate the energy associated with spin-spin interaction, and that of the SH beam is to capture the four-body interaction among spins. Incorporating all three, a Hamiltonian describing the chemical potential, two-body interaction, and four-body interaction can be effectively constructed. 

To derive the effective Hamiltonian, we consider a Gaussian input pump beam of wavelength $\lambda_p$, peak amplitude $E_0$, and beam waist $\mathrm{w}_{p}$. It shines a SLM whose phase mask consists of pixels ($m,n$) centered around $(x'_m,y'_n)$, each giving $0$ or $\pi$ phase modulation. The transverse electric field immediately after the SLM is approximately \cite{Pierangeli2019}
\begin{equation}
  E_p(x',y')=\sum_{m=1}^N \sum_{n=1}^N  \xi_{mn}\sigma_{mn}\frac{1}{a^2}\Pi\left(\frac{x'-x'_m}{a}\right)\Pi\left(\frac{y'-y'_n}{a}\right).
 \label{one}
 \end{equation}
Here $\xi_{mn}= E_0\exp\left[-(x'^2_m+y'^2_n)/\mathrm{w}_p^2\right]$ is the amplitude at pixel ($m,n$), $\Pi$ is the rectangular function of width $a$, and $\sigma_{mn}=\pm 1$ for the $0/\pi$ binary phase modulation.

The electric field is then transformed using a Fourier lens of focal length $F$ and coupled into a periodic-poled lithium niobate (PPLN) crystal of length $L$, so that it reads at the center of the crystal ($z=0$):  
\begin{equation}
 E_p(x,y,z=0)=\sum_{m=1}^N \sum_{n=1}^N \xi_{mn} \sigma_{mn}\eta_{mn} \mathrm{sinc}\left(\frac{a x \pi}{\lambda_p F}\right) \mathrm{sinc}\left(\frac{a y \pi}{\lambda_p F}\right) \exp(i\kappa_{p} z).
\end{equation}
Here, $\eta_{mn}=\exp\left[-2 \pi i (xx'_m+y y'_n)/\lambda_p F\right]$, $\kappa_{p}=(2\pi n_p)/\lambda_p$, and $n_p$ is the refractive index of the pump in the PPLN crystal. For simplification, we introduce contracted notations $\xi_i$ and $\sigma_i$, with $\xi_{i=m+(n-1)N }\equiv \xi_{mn}$, and $\sigma_{i=m+(n-1)N}\equiv \sigma_{mn}$, with $i=1,2,...N^2$ to index the $N \times N$ spins (pixels). In our setup, only near-axis light is fiber coupled and measured, so that $\mathrm{sinc}(a x \pi/\lambda_p F)$, $\mathrm{sinc}(a y \pi/\lambda_p F) \approx 1$, giving
\begin{equation}
 E_p(x,y,z)\approx\sum_{i=1}^{N^2} \xi_{i} \sigma_{i}\eta_i \exp(i \kappa_p z).
 \label{Eq_p}
 \end{equation}
In the PPLN crystal, the pump creates its SH according to the following dynamics,
\begin{equation}
2i\kappa_{p}\frac{\partial E_p}{\partial z}+\left(\frac{\partial^2}{\partial x^{2}}+\frac{\partial^2}{\partial y^2}\right)E_p=-2\frac{\omega_{p}^{2}}{c^{2}}\chi^{(2)}E_{p}^{*}E_he^{i\triangle\kappa z},\label{eq1}
\end{equation}
\begin{equation}
2i\kappa_h\frac{\partial E_h}{\partial z} +\left(\frac{\partial^2}{\partial x^{2}}+\frac{\partial^2}{\partial y^2}\right)E_h=-\frac{\omega_h^{2}}{c^{2}}\chi^{(2)}E_{p}^2 e^{-i\triangle\kappa z},
\label{eq2}
\end{equation}
which evolves from $-L/2$ to $L/2$. Here, $\kappa_h=(2\pi n_h)/\lambda_h$ is the wave number of the SH wave in the crystal with refractive index $n_h$. $\omega_{p}$ and $\omega_h$ are the angular frequencies of the pump and SH waves, respectively. $\Delta\kappa=2\kappa_{p}-\kappa_h-2\pi/\Lambda$ is the phase mismatching, with $\Lambda$ being the poling period.
Equations (\ref{eq1})--(\ref{eq2}) can in principle be solved by using split-step Fourier and adaptive step-size methods. However, the numeric solutions consume significant computational time and resources that increase exponentially with the spin number \cite{zhang_mode_2019}, whose inefficiency calls for the present optical realization. Only under the conditions of phase matching ($\Delta\kappa=0$), undepleted pump approximation, and negligibly small diffraction terms in Eqs.~(\ref{eq1}) and (\ref{eq2}), the transverse electric-field of the output SH wave can be obtained analytically as
\begin{equation}
 E_h(x,y,L/2)= i\frac{\omega_h^2\chi^{(2)}L}{c^2\kappa_h} E_p^2(x,y,-L/2). 
  \label{Eq_h}
 \end{equation}

At the crystal output, the pump and SH waves are each coupled into a single-mode fiber for detection, whose optical power is given by
\begin{equation}
P_{p,h} =\abs{\iint E_{p,h}(x,y) E^{p,h}_{f} dx dy }^2.  
\label{Eq_I}
\end{equation}
where \begin{equation}
E_{f}^{p,h}=\sqrt{\frac{2}{\pi}}\frac{1}{\mathrm{w}_{f}^{p,h}}\exp\left(-\frac{x^2+y^2}{(\mathrm{w}_{f}^{p,h})^2}\right), 
\label{Eq_f}
\end{equation}
are the normalized back-propagated fiber modes of beam waist $\mathrm{w}_{f}^{p}$ and $\mathrm{w}_{f}^{h}$ for the pump and SH waves, respectively. 

Substituting Eq.~(\ref{Eq_p}), (\ref{Eq_h}) and (\ref{Eq_f}) in Eq.~(\ref{Eq_I}), the detected power for the pump and SH waves is given in the form of
\begin{equation}
    P_{p}= \sum_{i=1}^{N^2} \sum_{j=1}^{N^2} J_{ij} \sigma_{i}\sigma_{j},
    \label{equ_P_det}
\end{equation}
and
\begin{equation}
    P_{h}= \sum_{i=1}^{N^2} \sum_{j=1}^{N^2} \sum_{s=1}^{N^2} \sum_{r=1}^{N^2}  J_{ijsr} \sigma_{i}\sigma_{j}\sigma_{s}\sigma_{r},\label{equ_h-det}
\end{equation}
where $J_{ij}$ and $J_{ijsr}$ prescribe the strength of the two-body and four-body interactions, respectively. As an example, Section 1 of the Supplementary Material presents the analytic results of $J_{ij}$ and $J_{ijsr}$ under the approximation of Eq.~(\ref{Eq_h}).

With Eq.~(\ref{equ_P_det}) and (\ref{equ_h-det}), we can define a single parameter $E$ to characterize the ``energy'' of the whole system, as
\begin{equation}
    E=\alpha C+\beta P_{p}+\gamma P_{h},
    \label{Eqn_E}
\end{equation}
where $C=\sum_{i=1}^{N^2} \mu_i \sigma_i$ is the weighted sum of spins that represents their total chemical energy, with the local chemical potential $\mu_i\in[-1,1]$. In this equation, $\alpha$, $\beta$, and $\gamma$ are free parameters defining the contribution of the chemical potential, two-body, and four-body interaction energy, respectively, to the total energy.

The total energy $E$ can be minimized by optimizing the binary phase mask on the SLM through adaptive feedback control (see Fig.~4). This is equivalent to finding the ground-state solutions of the effective Hamiltonian \begin{equation}
{\bf H} = \alpha \hat{H}_1 +\beta \hat{H}_2 + \gamma \hat{H}_4,
 \label{Eqn_H}
\end{equation}
where $\hat{H}_1$ is the Hamiltonian describing the chemical potential, $\hat{H}_2$ for the two-body and $\hat{H}_4$ for the four-body interaction, respectively, with  
\begin{equation}
\hat{H}_1= \sum_{i=1}^{N^2} \mu_i \hat{S}_i,
 \label{Eqn_H1}
\end{equation}
\begin{equation}
\hat{H}_2=\sum_{i=1}^{N^2} \sum_{j=1}^{N^2} J_{ij} \hat{S}_i \hat{S}_j,
\end{equation}
and
\begin{equation}
\hat{H}_4=\sum_{i=1}^{N^2} \sum_{j=1}^{N^2}\sum_{s=1}^{N^2} \sum_{r=1}^{N^2} J_{ijsr} \hat{S}_i \hat{S}_j \hat{S}_s \hat{S}_r. 
\end{equation}

\begin{figure*}[htbp]
    \centering
    \includegraphics[width=12cm]{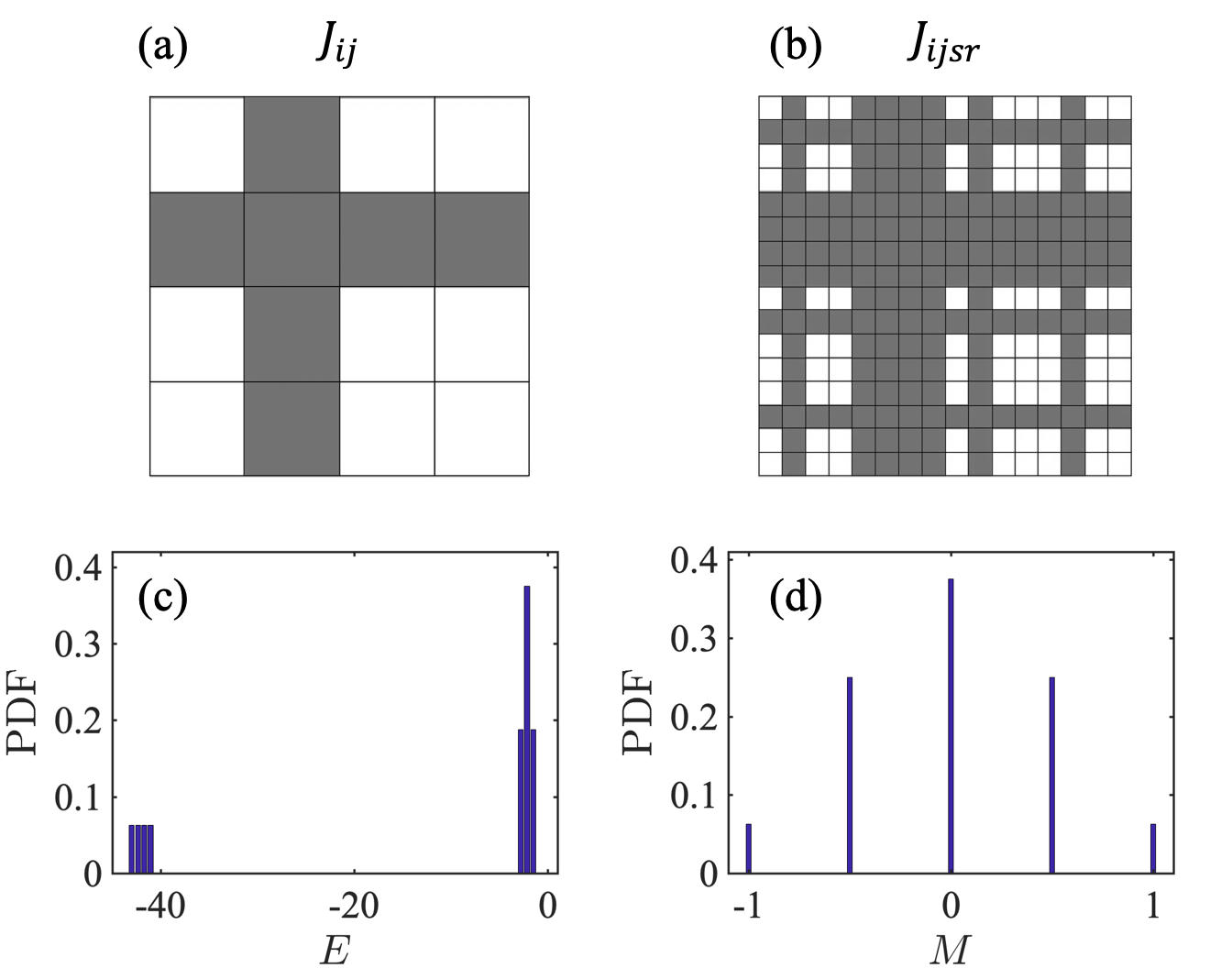}
    \caption{A toy Ising model with 4 spins, where the second spin $\sigma_2$ does not interact with the rest by blocking the input pump wavelet on the corresponding pixel. (a) shows the two-body interaction term $J_{ij}$, with $J_{i2}=J_{2j}=0$ for any $i,j$ (grey squares) and all other $J_{ij}=1$  (white squares).  (b) shows the four-body interaction term $J_{ijsr}$ by the same color scheme. 
    (c) and (d) plot the PDF of the energy and magnetization, respectively.}
    \label{example-4spin}
\end{figure*}

The two-body and four-body interactions can be tailored by modulating the input pump wave, varying the fiber optical modes, and modifying the phase matching conditions for the nonlinear process. As an example, Fig.~\ref{example-4spin} considers a toy system of four spins, with the input pump partially blocked and the output pump and SH light across the remaining pixels coupled equally into the fiber mode. The resulting interaction coefficients $J_{ij}$ and $J_{ijsr}$ are shown in Fig.~\ref{example-4spin}(a) and (b). 
For $\alpha=\beta=\gamma=-1$, the system's ground state is simply with all spin up. For all the possible spin configurations, we plot the probability distribution function (PDF) of the energy ($E$) and magnetization, defined as $M = \sum_{i=1}^{N^2} \sigma_i/N^2$, in Fig.~\ref{example-4spin} (c) and (d), respectively. As shown, the many-body interactions can be tailored into complex forms by simple linear optics operations. 


\section{\label{sec:level3}Experimental Setup}

 \begin{figure*}[htbp]
\centering 
\includegraphics[width=\linewidth]{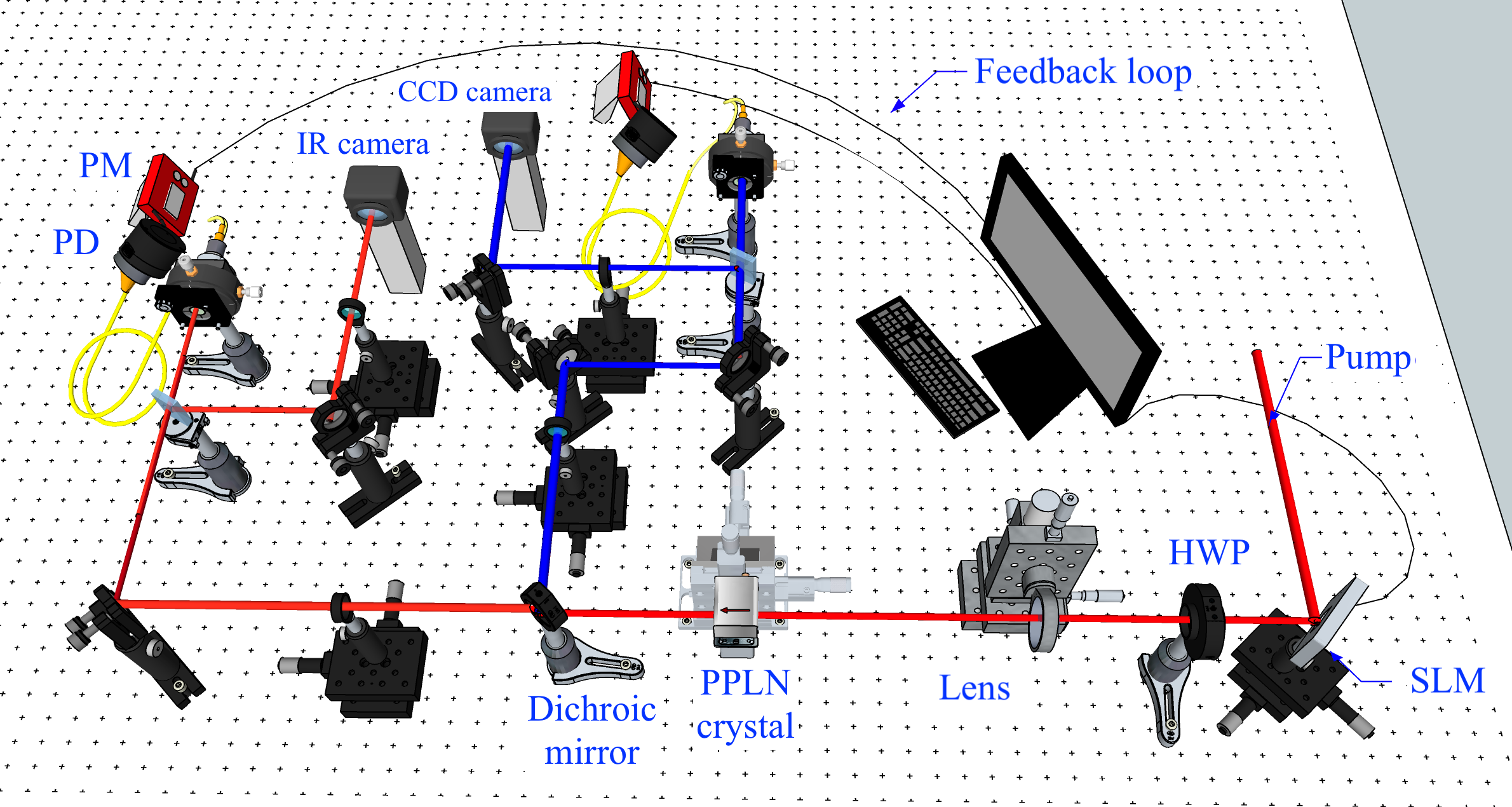}
\caption{ Experimental setup for the present nonlinear optical Ising machine. Pump laser pulses at wavelength 1551.5 nm with 5 ps pulse width and 50 MHz repetition rate are incident on a SLM, and focused into a PPLN crystal to generate SH light at wavelength 775.75 nm. After the crystal, the pump and SH beams are coupled into separate optial fibers and measured using power meters. SLM: Spatial Light Modulator, PPLN crystal: Magnesium-doped Periodic Poled Lithium Niobate crystal, PD: Photodiode, PM: Power meter, IR: Infrared, CCD: Charged Coupled Device and HWP: Half waveplate.  
} \label{ExpSetUp}
\end{figure*}

The experimental setup for the present nonlinear optical Ising machine is shown in Fig. \ref{ExpSetUp}. We use an optical pulse train at 1551.5 nm as the pump. Each pulse has 5 ps full-width at half-maximum (FWHM) and 50 MHz repetition rate. The pump's average power is about 40 mW and its pulse energy is $\sim$ 0.8 nanojoules. The transverse FWHM of the pump beam is 2.6 mm incident on the SLM (Santec SLM-100, 1440 $\times$ 1050 pixels, pixel pitch 10.4 $\times$ 10.4 $\mu$m) at a $50^{\circ}$ incidence angle \cite{Santosh19}. Initially, a random binary phase mask with phase value 0 or $\pi$ is uploaded onto the SLM. A lens focused the beams (focus length $F = $200 mm) inside a temperature-stabilized PPLN crystal with a poling period of 19.36 $\mu$m (5 mol.\% MgO doped PPLN, 10 mm length, 3 mm width, and 1 mm height from HC Photonics) for SH generation. The pump beam waist inside the crystal is 45 $\mu$m. The output is then filtered with a dichroic mirror to separate the SH and pump light \cite{QPMS2017}. Each arm is coupled into a single-mode fiber (SMF-28) using a fiber collimator consisting of aspheric lens (Thorlabs C220TMD-C and A375TM-B) and then detected by power meters (Thorlabs PM-100D with sensors S132C and S130C). The measurement results are sent to a computer through a MATLAB interface for feedback control, which updates the phase mask on the SLM to find the ground state of the customized Ising Hamiltonian. 

\section{Results and discussions}

\begin{figure*}[htbp]
\centering 
\includegraphics[width=10cm]{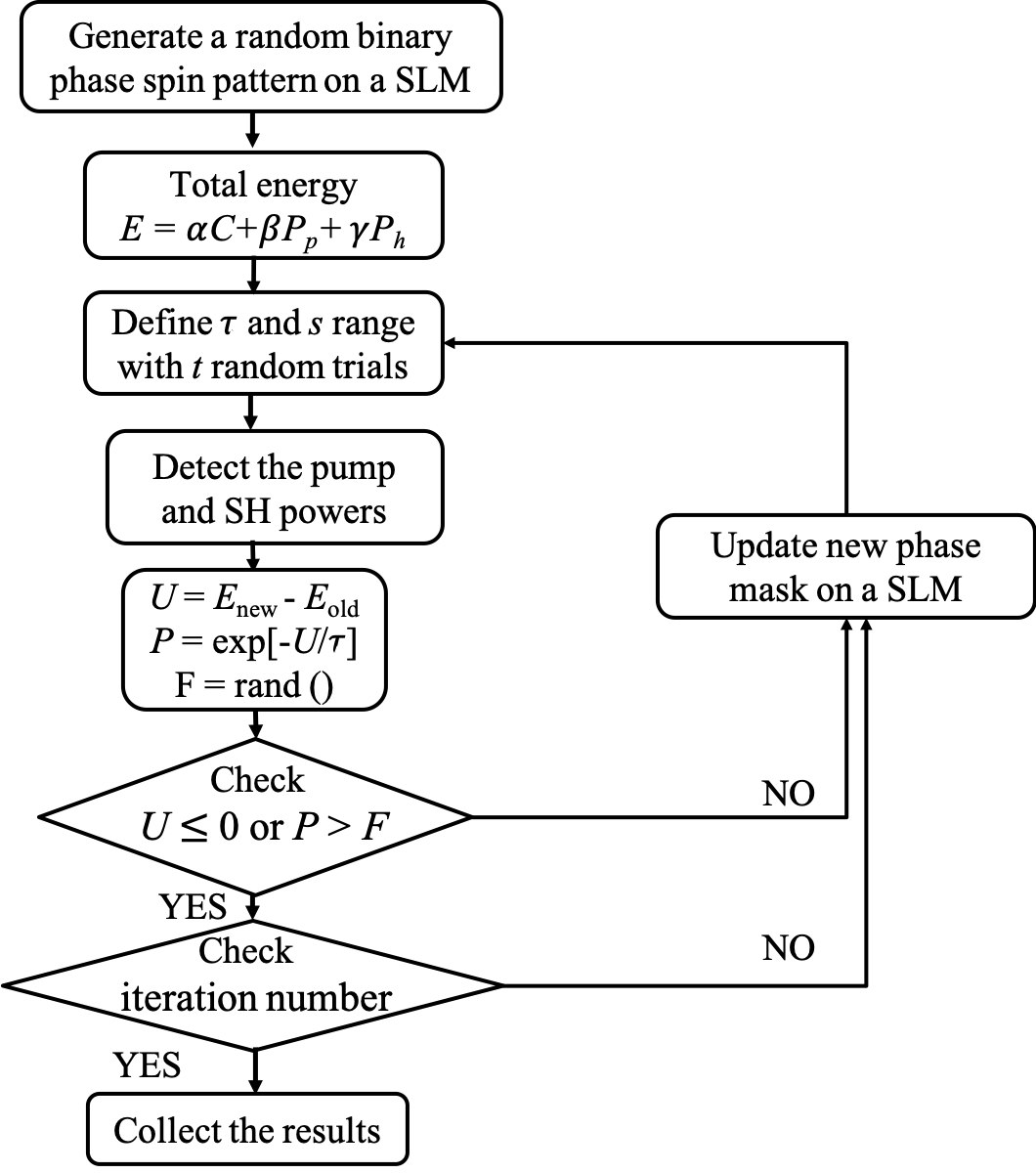}
\caption{A flow chart of the adaptive feedback control algorithm using a Monte Carlo method. Step 1: generate an initial random binary phase spin pattern on a SLM. Step 2: define the total energy of the system. Step 3: define the range for $s$ and $\tau$ and run $t$ random trials for each. Step 4: detect the pump and SH powers. Step 5: find the energy difference $U=E_{new}-E_{old}$ (where $E_{old}$ is the previous minimum energy), Boltzmann's probability function $P=\exp[-U/\tau]$ and a random variable $F\in [0,1]$. Step 6: check if the optimization criteria, $U\leq$ 0 or $P$ $>$ $F$, is met. Step 7: check the number of iterations. Step 8: if the criteria is not satisfied in Step 6 and Step 7, flip the spins within a randomly chosen cluster and update the binary phase mask on a SLM. Step 9: repeat Steps 3-8 if optimum criteria is not satisfied. Step 10: stop the feedback loop and collect the results.}
\label{flow_chart}
\end{figure*}

As shown in Eq.~(\ref{Eqn_H1}), the chemical potential of each spin is flexibly defined by $\mu_i$ and their collective contribution to the total energy is controlled by $\alpha$. This provides the knob of studying the magnetization under a variety of local and global single-spin parameters. For this paper, however, we will focus on the many-body interaction and consider only $\mu_i=1$ in all of the following results. Meanwhile, we will leave fine tuning two and four-body interaction to our future work, but only control each's aggregated contribution to the total energy by varying $\beta$ and $\gamma$, respectively. 

To find the ground states of the total Hamiltonian, the SLM's initial phase mask is prepared in small clusters with randomly chosen 0 or $\pi$ phases. The resulting pump and SH waves are coupled separately into single mode fibers, whose optical power is measured for feedback control. To minimize the total energy, we adaptively flip the spins within a randomly chosen cluster, following the standard Monte Carlo approach \cite{landau_guide_2014}. A flow chart of this procedure is shown in Fig.~\ref{flow_chart}, where the spin flipping during each iteration is accepted or rejected according to a Boltzmann's probability function $P=\exp[-U/\tau]$, with $U = E_{new}-E_{old}$ being the change in energy and $\tau$ the thermal energy. To avoid trapping into a local minimum, we vary both the cluster size $s$ and $\tau$ during iterations. Note that this algorithm is not necessarily the most efficient, but nevertheless adequate for our current demonstrations as the first case study in this new Ising platform. A machine-learning based Monte Carlo method could be utilized in the future to speedup this optimization  \cite{liu_self-learning_2017,shen_self-learning_2018}. 

\begin{figure*}[htbp]
\centering 
\includegraphics[width=8.1cm]{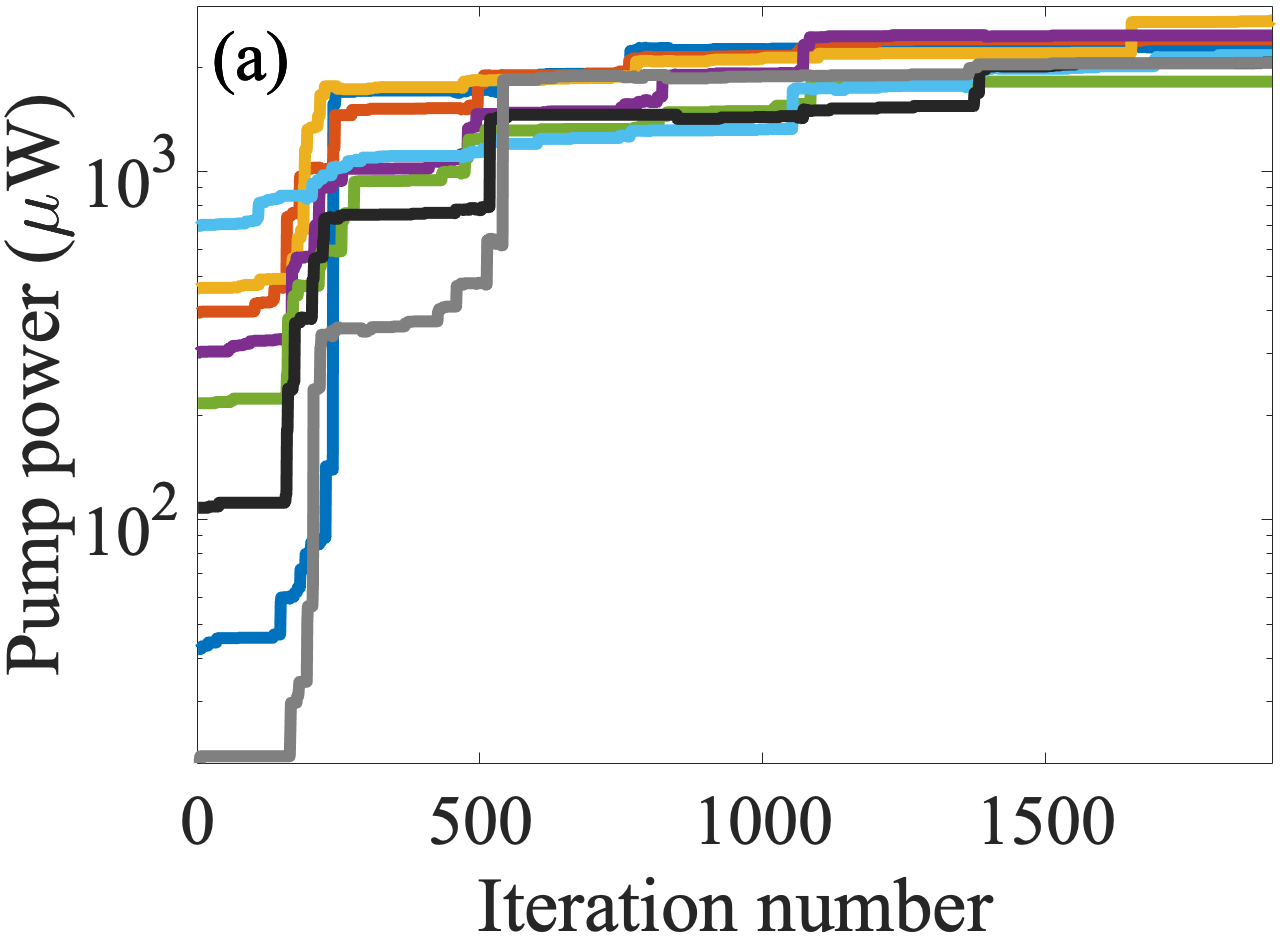}
\includegraphics[width=8.1cm]{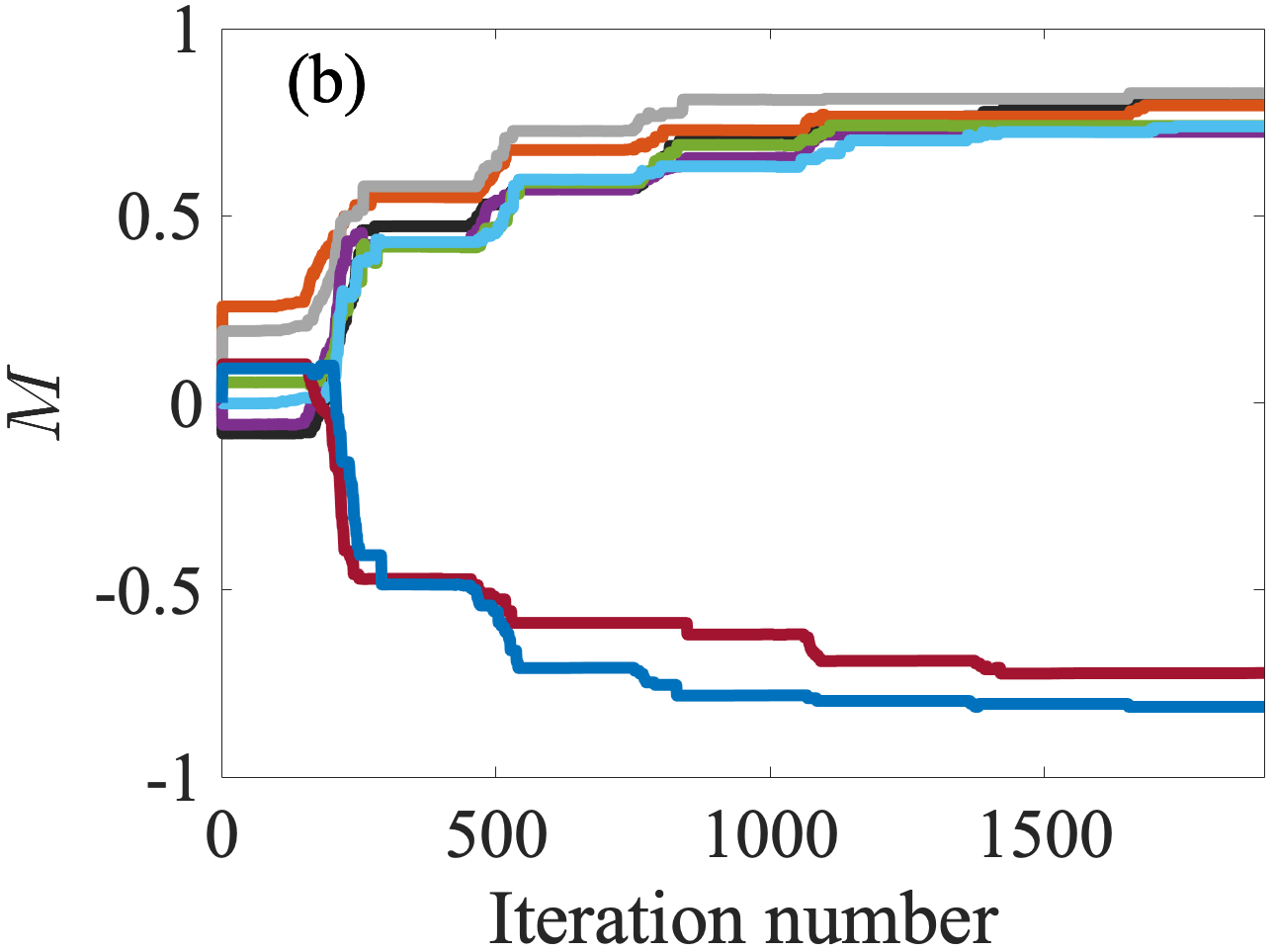}
\caption{ Measured pump power (a) and magnetization (b) over optimization iterations for 800$\times$800 spins with $\alpha=0$, $\beta=-1$, and $\gamma=0$. 
} 
\label{PDF_Results_1}
\end{figure*}

 
Figure~\ref{PDF_Results_1} illustrates the process of optimization for 800$\times$800 spins with eight initial random phase masks. Figure~\ref{PDF_Results_1} (a) shows how the optical pump power is increased, thus the decrease of total energy $E$ to approach the ground state of the system. With $\alpha=0$, there is spontaneous symmetry breaking, as the system energy remains unchanged if all spins are flipped. As such, the feedback control will optimize the spins toward either positive or negative magnetic states with equal probability \cite{Pierangeli2019}. This is clear in Fig.~\ref{PDF_Results_1} (b), where the magnetization trends both ways. For all initial phase trials, the absolute value of average magnetization can reached to $\sim$ 0.75. 


\begin{figure*}[htbp]
\centering 
\includegraphics[width=8.15cm]{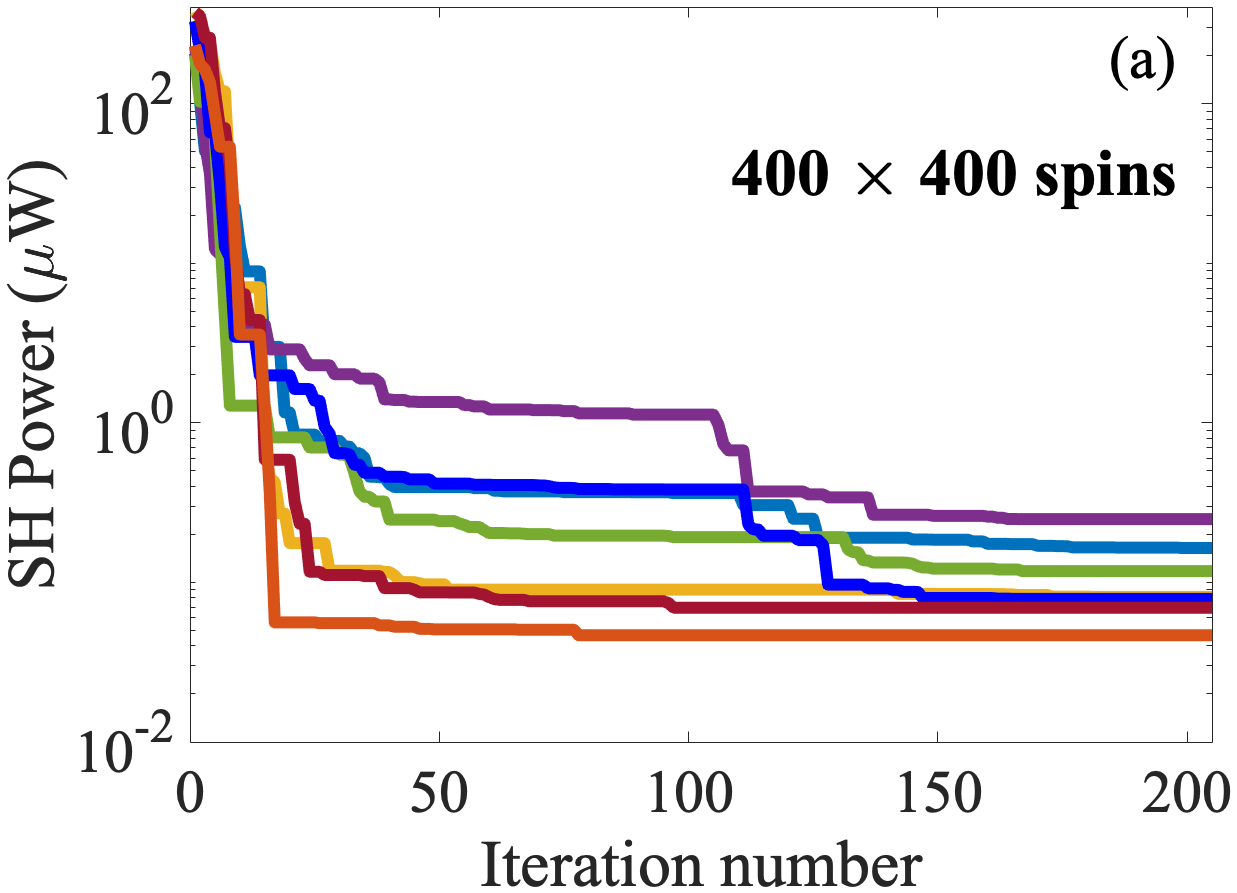}
\includegraphics[width=8.15cm]{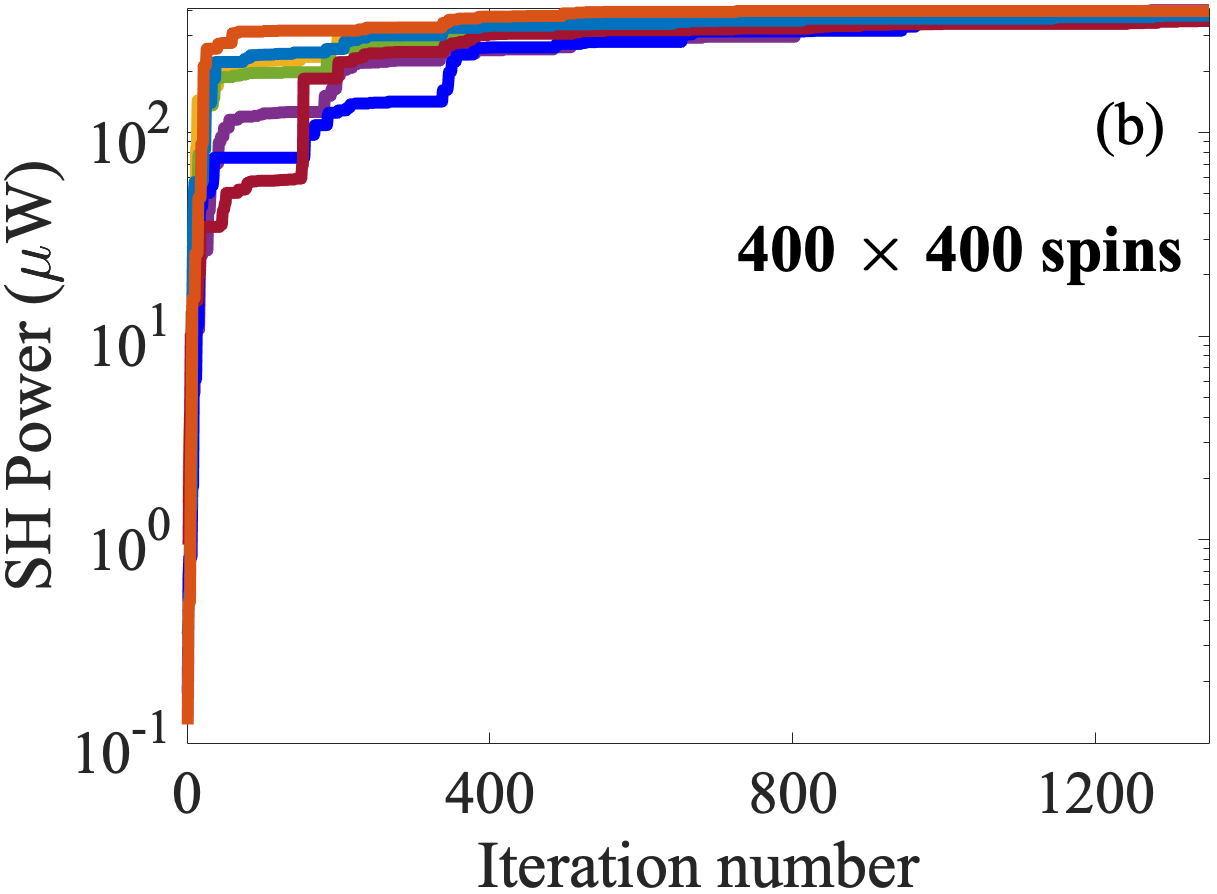}
\includegraphics[width=8.15cm]{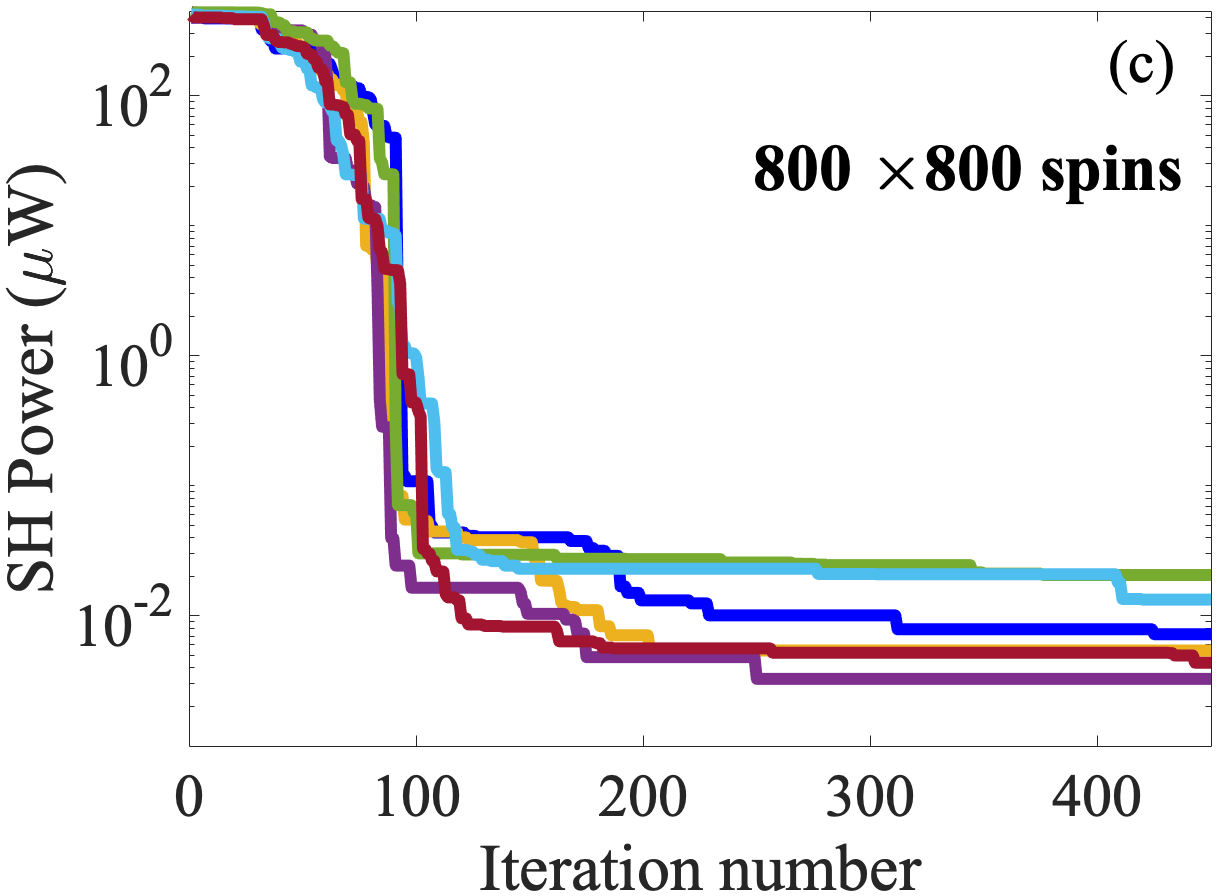}
\includegraphics[width=8.15cm]{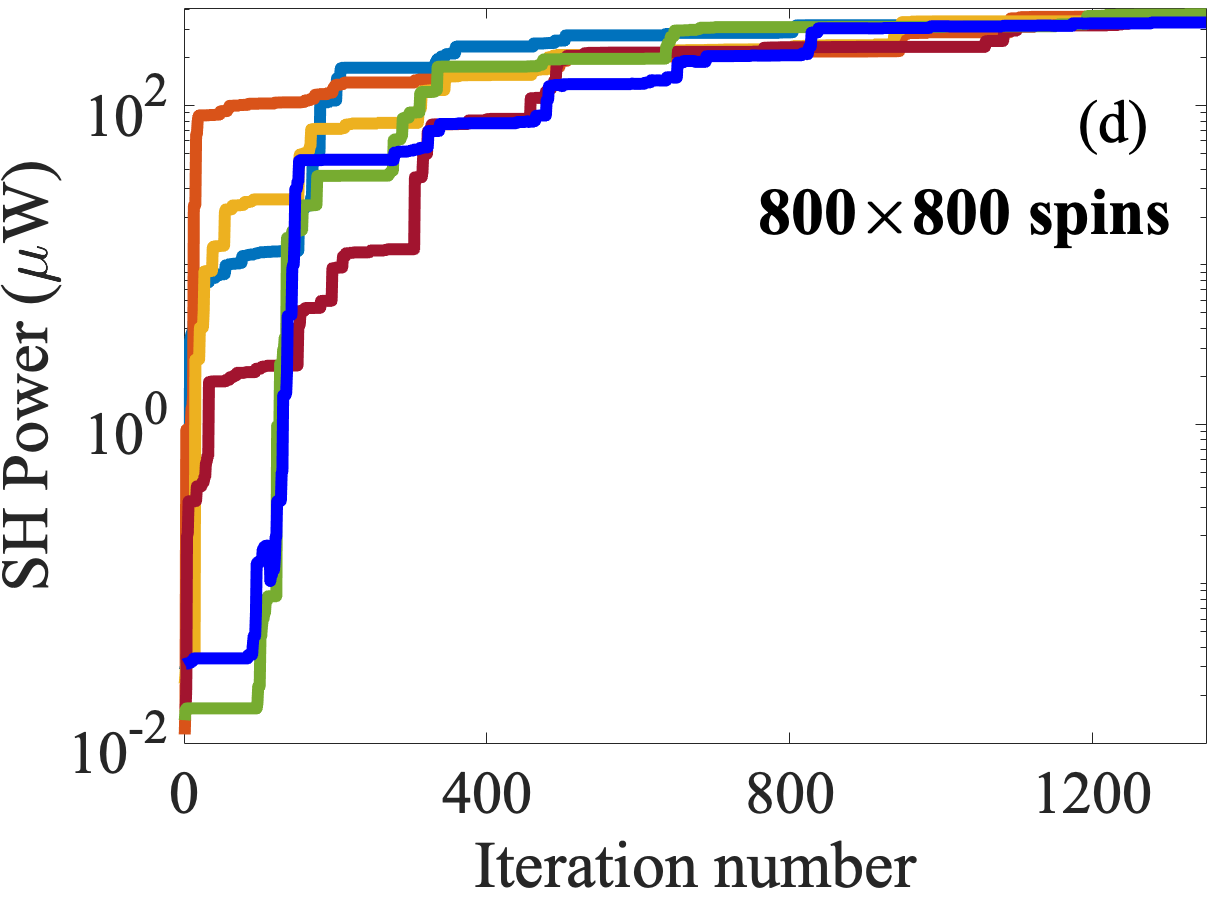}
\caption{Measured SH power (in log scale) evolving during the feedback-control optimization. (a) shows the decrease in power over iterations for 400$\times$400 spins and an Ising machine with $\alpha=0$, $\beta=0$, and $\gamma=1$. 
(b) shows the increase in SH power, which corresponds to approximate the ground state of an Ising machine with $\alpha=0$, $\beta=0$, and $\gamma=-1$. (c) and (d) show similar results but for 800$\times$800 spins. Here, curves in different colors are for the results obtained from different initial random phase masks.} 
\label{400n800spins}
\end{figure*}

\begin{figure*}[htbp]
\centering 
\includegraphics[width=\linewidth]{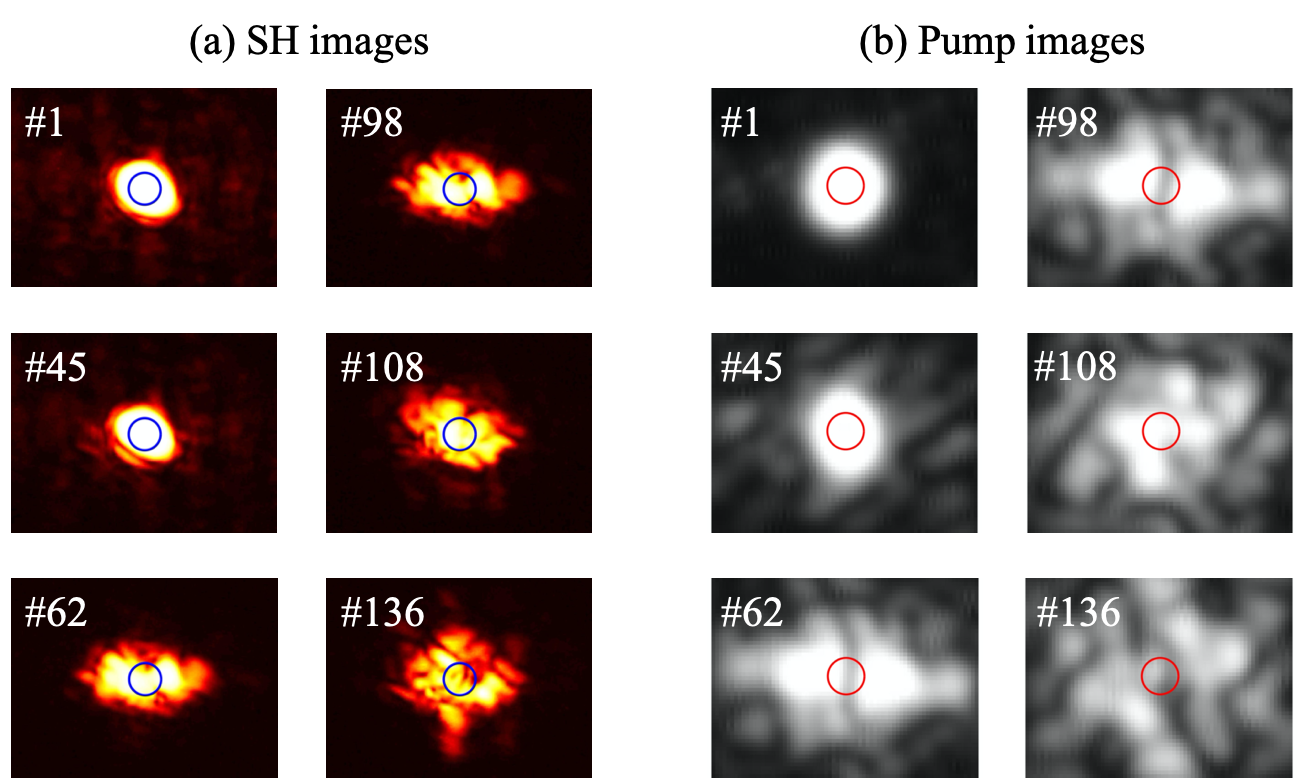}
\includegraphics[width=9cm]{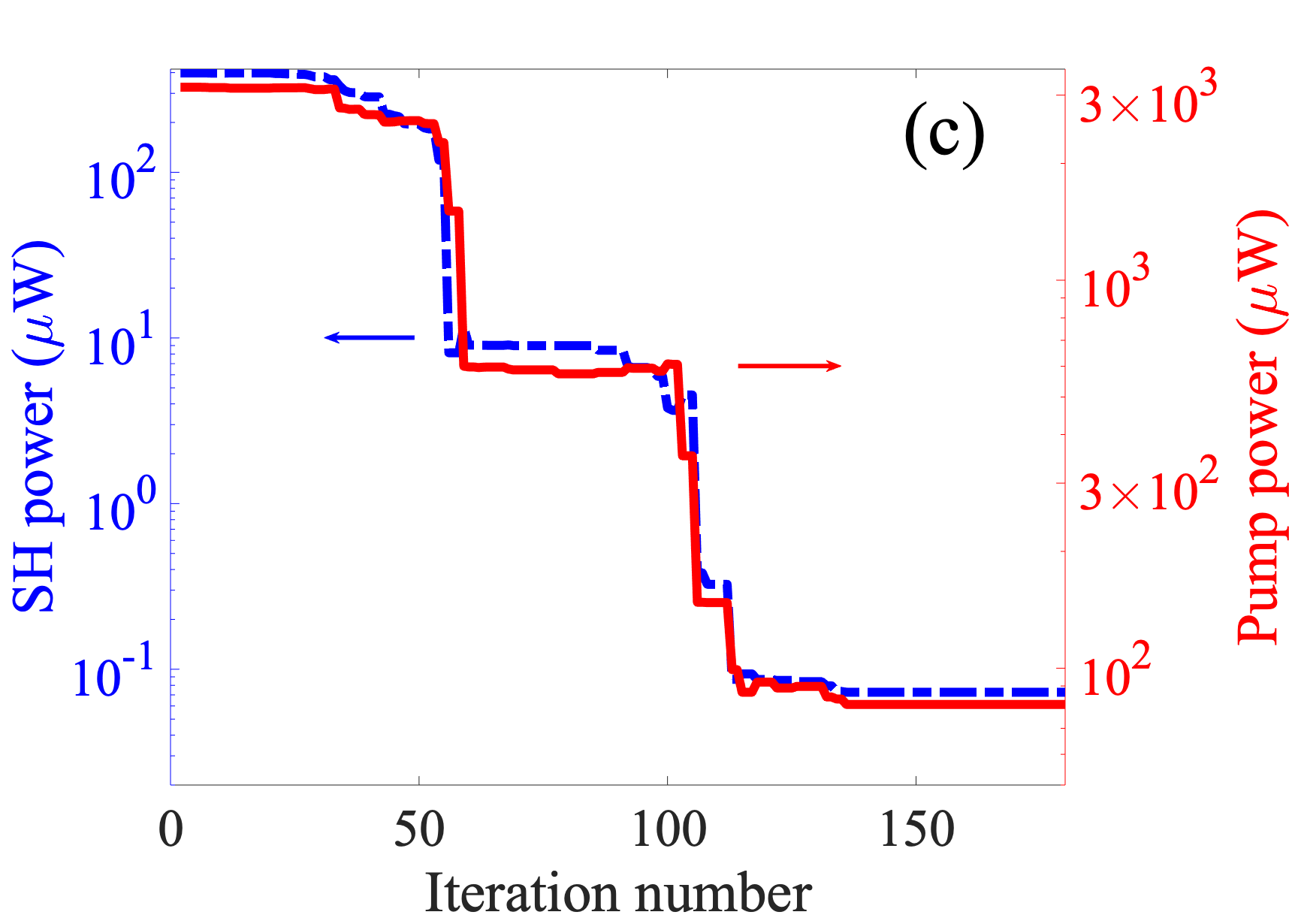}
\caption{(a) and (b): Resulting images of the SH and pump lights at the crystal output upon different numbers of iterations. The system contains 800$\times$800 spins and the goal is to find the ground state of the Ising Hamiltonian with $\alpha=0$, $\beta=1$, and $\gamma=1$. The blue and red circles indicate the coupling areas of the SH and pump light into the single mode fibers, respectively. (c) Shows the measured SH (blue dashed curve) and pump (red solid curve) power over iterations. 
} 
\label{Images_1}
\end{figure*}

In Fig.~\ref{400n800spins}, we show the two sets of measurements for $400\times400$ and $800\times800$ spins, respectively, with purely four-body interaction, i.e., $\alpha=0$ and $\beta=0$. (a) and (c) plot the SH power evolution for $\gamma=1$, for which the system ground states correspond to the minimum SH power. In both figures, different initial spins are optimized to give similar minimum SH power, which indicates the robustness of our optimization method. For $\gamma=1$, the system evolves into a paramagnetic-like state that minimizes the SH power in the fiber. In Fig. \ref{400n800spins}(c), the values are close to the minimum detectable power of our optical sensor ($\sim$5 nW). Because of a smaller pixel size in (c), the spin disorder is stronger to give lower SH power in (a). In opposite, (b) and (d) are for $\gamma=-1$, where the ground states are obtained at the highest SH power. In this case, the system exhibits a ferromagnetic-like behavior. For $400\times400$ and $800\times800$ spins, the optimization leads to similar maximum SH power despite different initial spin conditions. The convergence is slower for the latter case, as there are four times more spins to be optimized. Overall, our system can reliably and efficiently evolve into the vicinity of its ground state.
 

To further understand the optimization mechanism, we take images of the pump and SH beams at the crystal output by splitting them using flipping beam-splitters, as shown in Fig.~\ref{ExpSetUp}. Through 4f systems, the pump is imaged on a NIR-IR camera (FIND-R-SCOPE Model No. 85700 with pixel resolution of 17.6$\mu$m), and the SH on a CCD camera (Canon Rebel T6 with pixel pitch of 4.3$\mu$m). Figure \ref{Images_1} (a) and (b) show the pump and SH images, respectively, under different iteration numbers as they minimize the energy of 800$\times$800 spin system with $\alpha=0$, $\beta=1$, and $\gamma=1$. As shown, both light beams become scattering and show speckle patterns as the optimization goes. The resulting fiber-coupled SH and pump power is shown in Fig.~\ref{Images_1} (c). Both decrease with the iteration numbers, dropping by orders of magnitude to minimum values after 150 iterations. Note that the final SH power level is very close to the detection level of the sensors, which prohibits further reduction via the present feedback control.   

\begin{figure*}[htbp]
\centering 
\includegraphics[width=8.15cm]{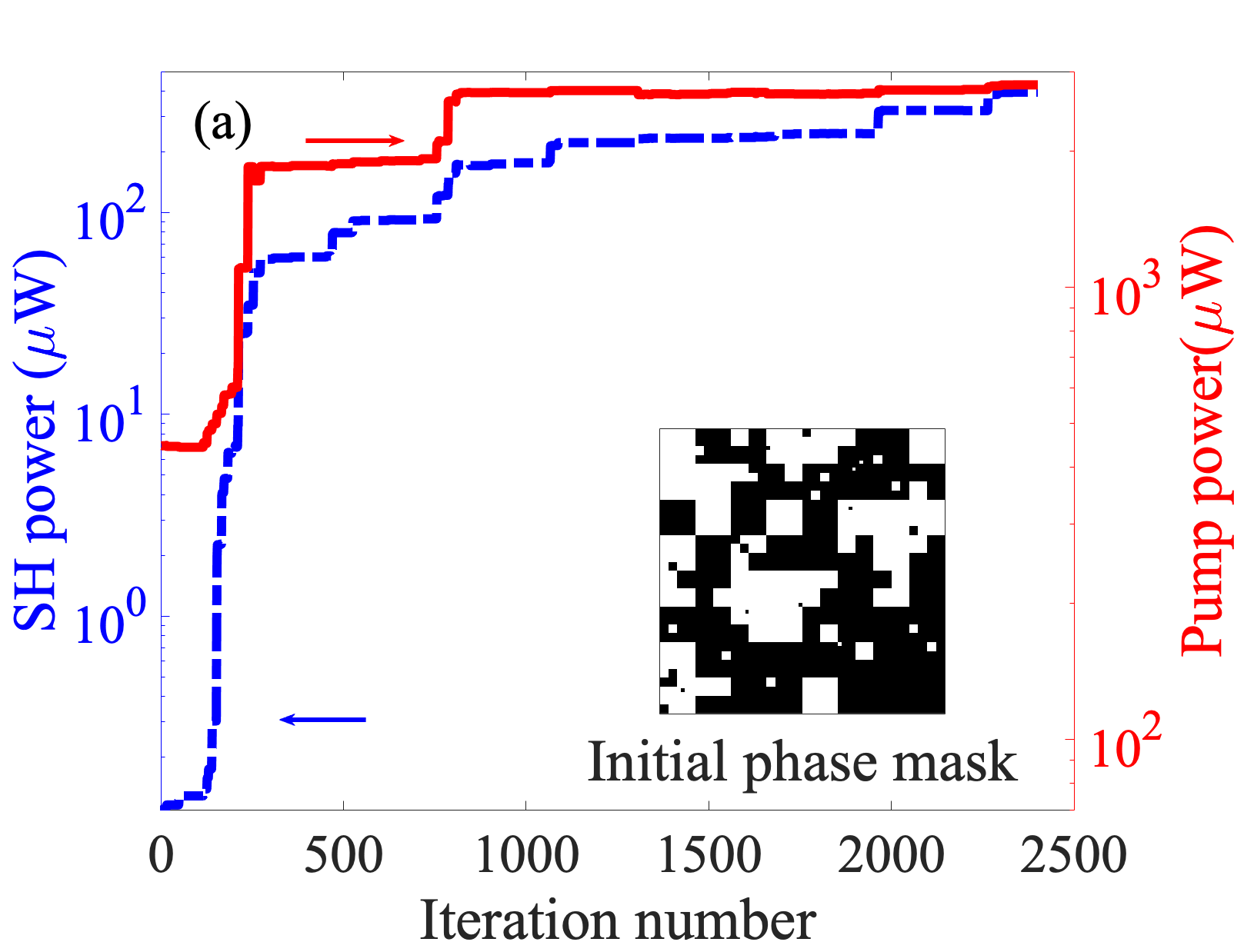}
\includegraphics[width=8.15cm]{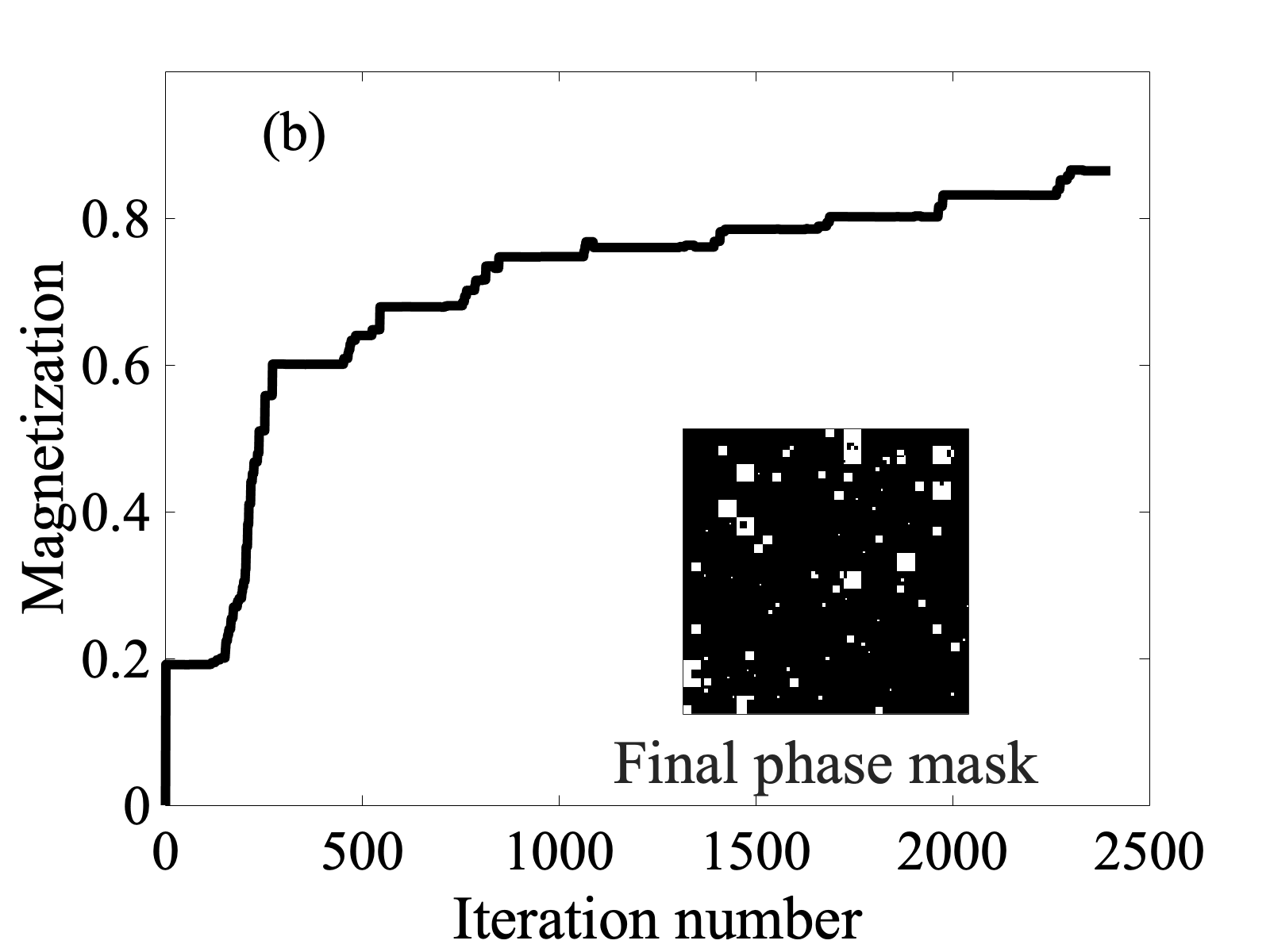}
\includegraphics[width=8.15cm]{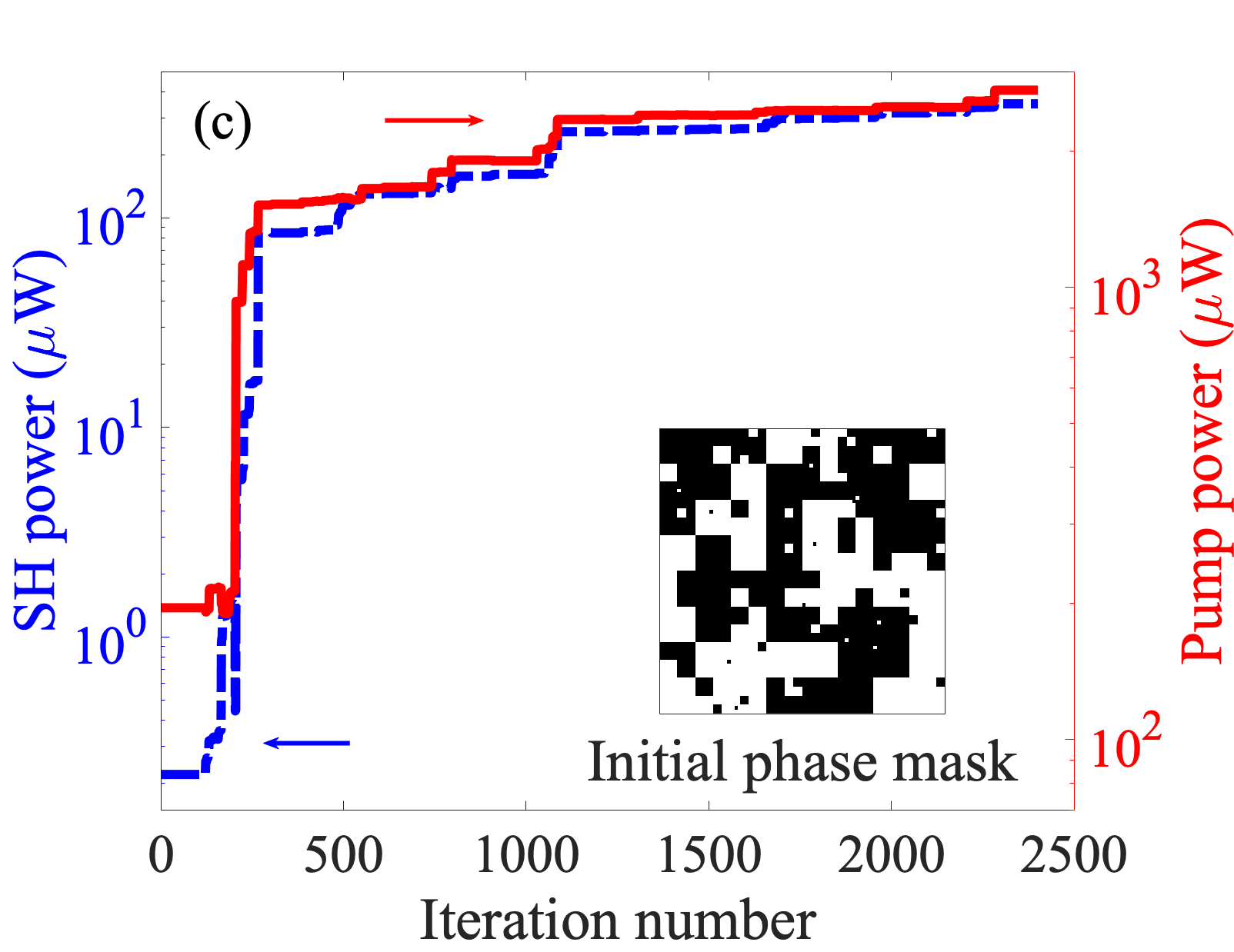}
\includegraphics[width=8.15cm]{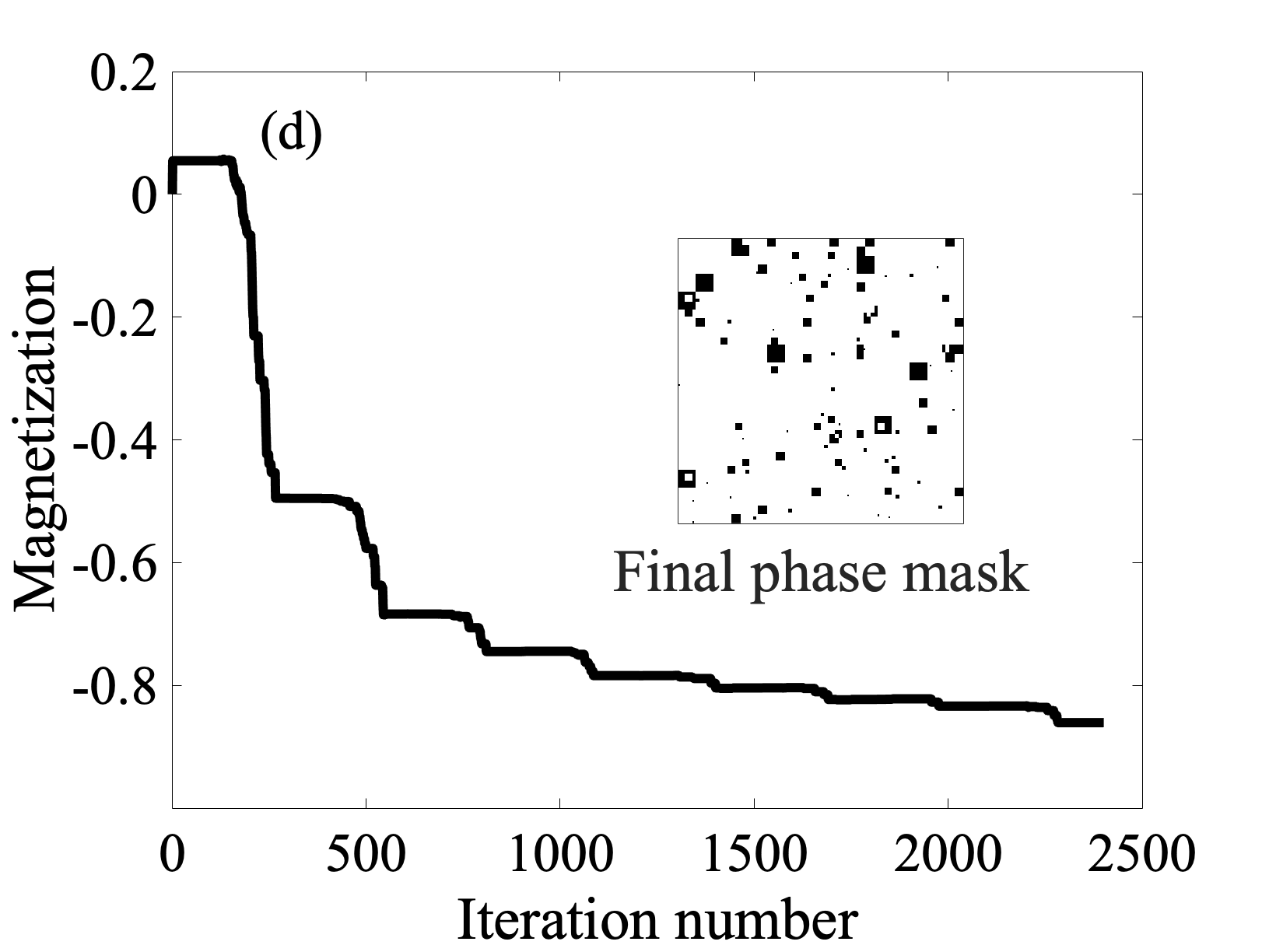}
\caption{ (a) Experimentally measured evolution of the SH power (blue dashed curve) and pump (red solid curve) power during optimization, set to find the ground state of the Ising machine with $\alpha=-1$, $\beta=-0.5$, and $\gamma=-1$. (b) shows the evolution of magnetization. (c) and (d) are similar results but with $\alpha=1$, $\beta=-0.5$, and $\gamma=-1$. Insets show the initial and final phase masks in each case, respectively, where the black and white pixels indicates the positive and negative orientated spins, respectively.} 
\label{pic_Results}
\end{figure*}

As seen in Eq.~(\ref{Eqn_H}), the two-body and four-body interaction energies are dependent only on the relative alignment of the spins, but not on each's absolute orientation. Thus spontaneous symmetry breaking could occur during the optimization, leading to bifurcation \cite{Pierangeli2019}. To avoid this symmetry breaking, one could set a non-zero $\alpha$ to control the convergence direction of spin optimization. As an example, in Fig. \ref{pic_Results}, we compare the results with $\alpha=1$ and $-1$, both with $\beta=-0.5$ and $\gamma=-1$. As shown, $\alpha$ can indeed dictate the spin alignment to result in either positive or negative magnetization states. In both cases, starting from a rather randomized phase mask, the spins become relatively aligned to increase the SH and pump power, but magnetization can have positive or negative orientation as can be seen Fig. \ref{pic_Results}(b) and (d). The inset of Fig. \ref{pic_Results}(b,d) shows the final phase mask of the optimum solution, where black and white colors represents the positive and negative orientation of the spins, respectively. Similar results are shown in Section 2 of the Supplementary Material for $\beta=-1$.

\begin{figure*}[htbp]
\centering 
\includegraphics[width=8.15cm]{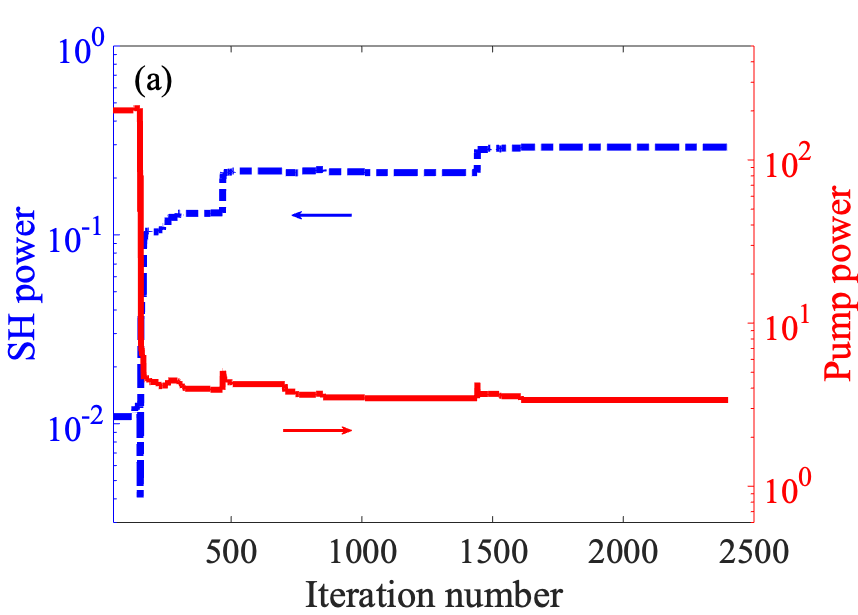}
\includegraphics[width=8.15cm]{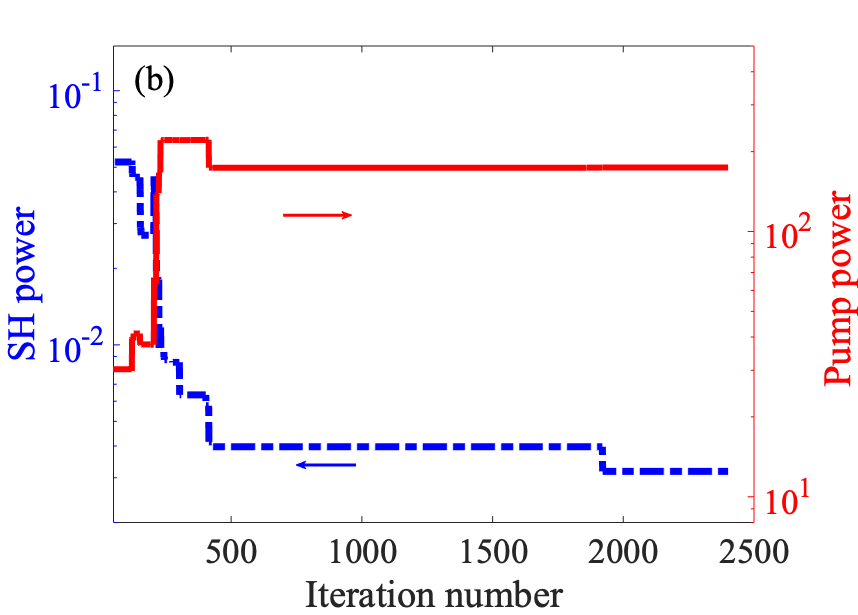}
\caption{Experimentally measured evolution of the SH (blue dashed curve) and pump (red solid curve) power during optimization for (a) $\alpha=0$, $\beta=1$, and $\gamma=-1$, and (b) $\alpha=0$, $\beta=-1$, and $\gamma=1$. 
} 
\label{pic_Resultsa}
\end{figure*}

We last consider the cases where two-body and four-body interactions contribute oppositely to the total energy of the system. For instance, the two-body interaction can be attractive but four-body be repulsive, or vice versa. Such systems can be conveniently configured by defining the pre-factors $\beta$ and $\gamma$ in Eq.~(\ref{Eqn_H}). The optimization will maximize one while minimizing the other. As an example,  Fig.~\ref{pic_Resultsa} plots the optimization trajectory for opposite two-body and four-body interactions. In Fig.~\ref{pic_Resultsa}(a), $\beta=1$, and $\gamma=-1$, so that the system energy is maximized by reducing the pump power while increasing the SH power. Its opposite configuration is in Fig.~\ref{pic_Resultsa}(b), where $\beta=-1$, and $\gamma=1$ so that the same optimization increases the pump power while reducing the SH power. This example suggests that the nature of two-body and four-body interactions can be conveniently maneuvered in our Ising machine, which makes it versatile for simulating various systems in solid state physics \cite{Oitmaa_1973,Landau83,PhysRevE.64.036126,Li_2017}, chemical engineering \cite{Keil:2011}, and so on. This Ising machine is efficient because it calculates the many-body interaction energy during a single pass through a nonlinear crystal, realizing simultaneously matrix multiplication, Fourier transformation, etc., of large-size data.

\section{Conclusion}
Using spatial phase modulation and second-harmonic generation, we have constructed Ising emulators with all-to-all connections and tailored chemical potential, two-body interaction, and four-body interaction. Their ground-state solutions can be effectively and reliably approximated by adaptive feedback control, whose speed is currently limited by the processing time of the spatial light modulator. A significant speedup is achievable by using ferroelectric liquid-crystal based spatial light modulators or programmable plasmonic phase modulators \cite{liu_focusing_2017,smolyaninov_programmable_2019}. At the present, the maximum number of accessible spins is about 1 million, and can be further increased by using modulators with more pixels or combining multiple modulators. While this study considers only second-harmonic generation, it can be straightforwardly extended to other nonlinear processes, such as sum-frequency generation, difference-frequency generation, four-wave mixing, for other interesting Ising machines \cite{Vogt1974,four-wave_mixing,QPMS2017,Santosh19,zhang_mode_2019}. Such nonlinear optical realizations of Ising machine could contribute as supplements for big data optimization and analyses that remain challenging for classical super-computers or forthcoming quantum machines but with a limited number of qubits.



\begin{addendum}
 \item[Competing Interests] The authors declare that they have no competing interests.
\end{addendum}

{\bf References }

\newpage

{\bf \Large  Supplementary Materials 
}

\beginsupplement



\section{Theoretical description for two-body and four-body interaction terms}
To present the analytic results of $J_{ij}$ and $J_{ijsr}$, Eq. (9) and (10) of the main text are described in this section. 
The detected pump power and SH power, each coupled into the single mode fiber, are given by
\begin{equation}
P_{p,h} =\abs{\iint E_{p,h}(x,y) E^{p,h}_{f} dx dy }^2, 
\label{Eq_I_s}
\end{equation}
respectively, where 
\begin{equation}
 E_p(x,y,z)\approx\sum_{i=1}^{N^2} \xi_{i} \sigma_{i}\eta_i \exp(i \kappa_p z),
 \label{Eq_p_s}
 \end{equation}
is the electric field of the pump wave,
 \begin{equation}
 E_h(x,y,L/2)= i\frac{\omega_h^2\chi^{(2)}L}{c^2\kappa_h} E_p^2(x,y,-L/2), 
  \label{Eq_h_s}
 \end{equation}
is the electric field of the SH wave under the condition of undepleted pump approximation with phase matching and negligibly small diffraction, and 
\begin{equation}
E_{f}^{p,h}=\sqrt{\frac{2}{\pi}}\frac{1}{\mathrm{w}_{f}^{p,h}}\exp\left(-\frac{x^2+y^2}{(\mathrm{w}_{f}^{p,h})^2}\right), 
\label{Eq_f_s}
\end{equation}
is the normalized back-propagated fiber modes for the pump and SH waves, respectively.
 After substitute Eq. (\ref{Eq_p_s}), (\ref{Eq_h_s}) and (\ref{Eq_f_s})  into Eq. (\ref{Eq_I_s}). The detected optical power of pump wave is 
\begin{equation}
    P_{p} = 2\pi (\mathrm{w}_{f}^{p})^2 \sum_{i=1}^{N^2} \sum_{j=1}^{N^2} \xi_{i}\xi_{j}\xi_{i}^{f}\xi_{j}^{f} \sigma_{i}\sigma_{j},
\end{equation}
and the detected optical power of SH wave is
\begin{equation}
    P_{h}= 2\pi (\mathrm{w}_{f}^{h})^2A^2\sum_{i=1}^{N^2} \sum_{j=1}^{N^2} \sum_{s=1}^{N^2} \sum_{r=1}^{N^2}  \xi_{i}\xi_{j}\xi_{s}\xi_{r}\xi_{is}^{'f}\xi_{jr}^{'f}\sigma_{i}\sigma_{j}\sigma_{s}\sigma_{r},
\end{equation}
where 
\begin{equation}
\xi_{i= \{m+(N-1)n\}}= E_0\exp\left[-(x'^2_m+y'^2_n)/\mathrm{w}_p^2\right],
\end{equation}
\begin{equation}
\xi_{i= \{m+(N-1)n\}}^f = \exp\left[- \frac{\pi^2(\mathrm{w}_{f}^{p})^2}{\lambda_p^2F^2}(x'^2_m+y'^2_n)\right],
\end{equation}
\begin{equation}
\xi_{i= \{m+(N-1)n\}s = \{l+(N-1)k\}}^{'f} = \exp\left[- \frac{\pi^2(\mathrm{w}_{f}^{h})^2}{\lambda_p^2F^2}\left( (x'_m+x'_l)^2+(y'_n+y'_k)^2\right)\right],
\end{equation}
and
$A= (i\omega_h^2\chi^{(2)}L) /c^2\kappa_h$. $\mathrm{w}_{f}^{p}$ and $\mathrm{w}_{f}^{h}$ are the beam waists of the normalized back-propagated fiber modes for pump and SH waves, respectively. It gives
\begin{equation}
    P_{p} = \sum_{i=1}^{N^2} \sum_{j=1}^{N^2} J_{ij} \sigma_{i}\sigma_{j},
    \label{equ_P_det1}
\end{equation}
and
\begin{equation}
    P_{h} = \sum_{i=1}^{N^2} \sum_{j=1}^{N^2} \sum_{s=1}^{N^2} \sum_{r=1}^{N^2}  J_{ijsr} \sigma_{i}\sigma_{j}\sigma_{s}\sigma_{r},\label{equ_h-det1}
\end{equation}
where $J_{ij} =  2\pi (\mathrm{w}_{f}^{p})^2\xi_{i}\xi_{j}\xi_{i}^{f}\xi_{j}^{f}$ and $J_{ijsr} = 2\pi (\mathrm{w}_{f}^{h})^2 A^2\xi_{i}\xi_{j}\xi_{s}\xi_{r}\xi_{is}^{'f}\xi_{jr}^{'f}$ are the two-body, and four-body interaction terms, respectively.

\section{Results for the evolution of the pump and its SH with magnetization}

\begin{figure}[htbp]
\centering 
\includegraphics[width=8.0cm]{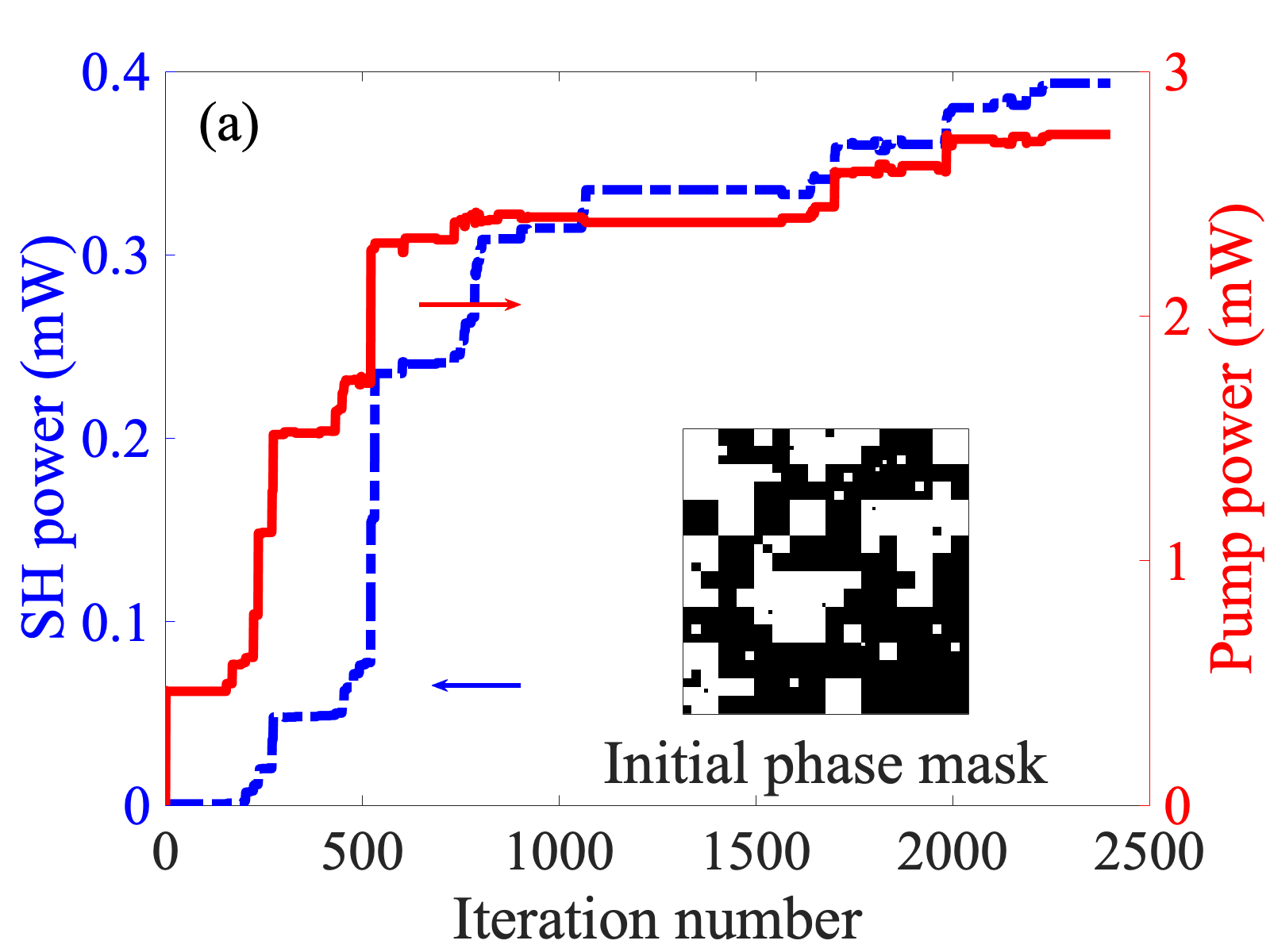}
\includegraphics[width=8.0cm]{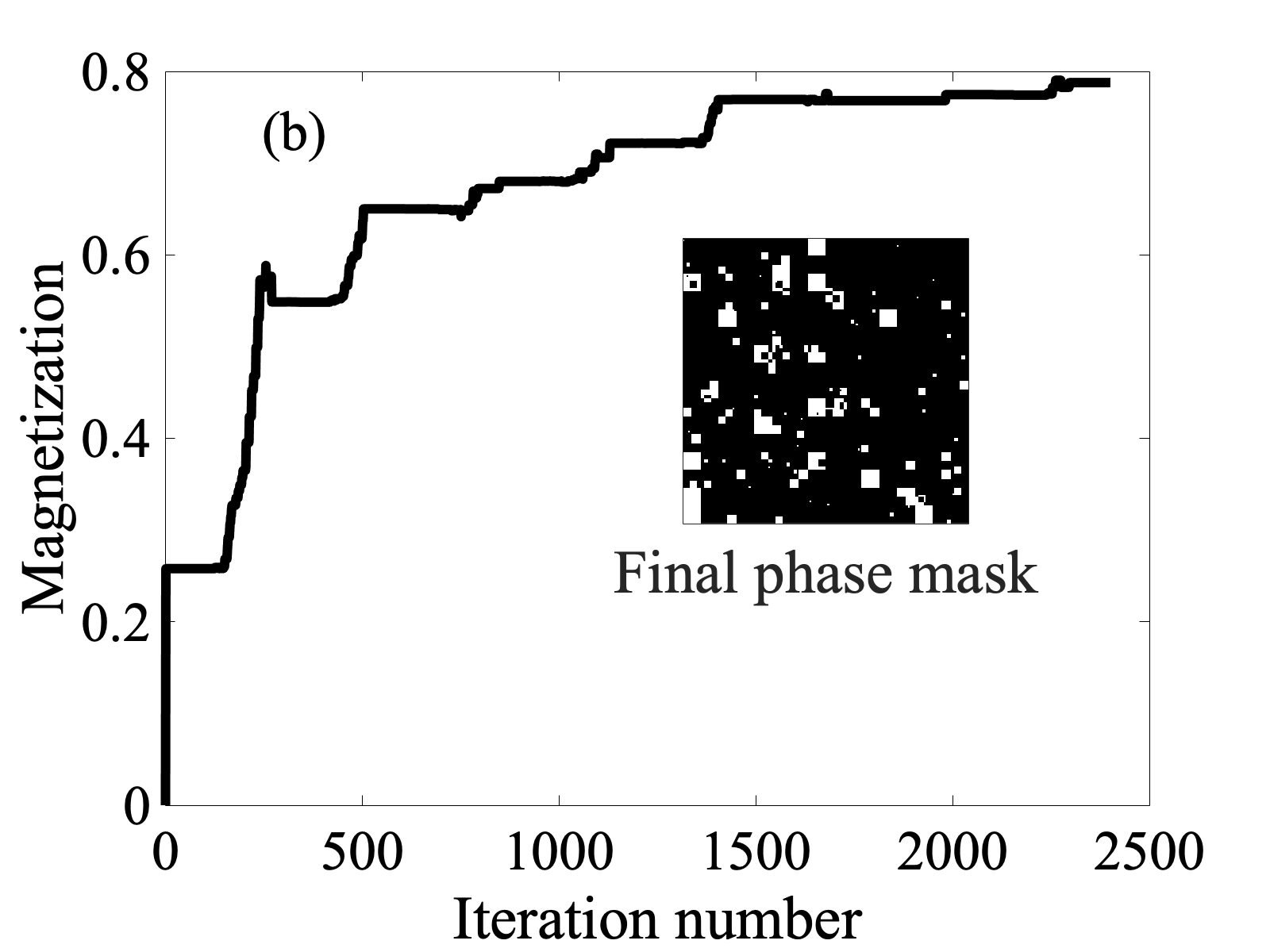}
\includegraphics[width=8.0cm]{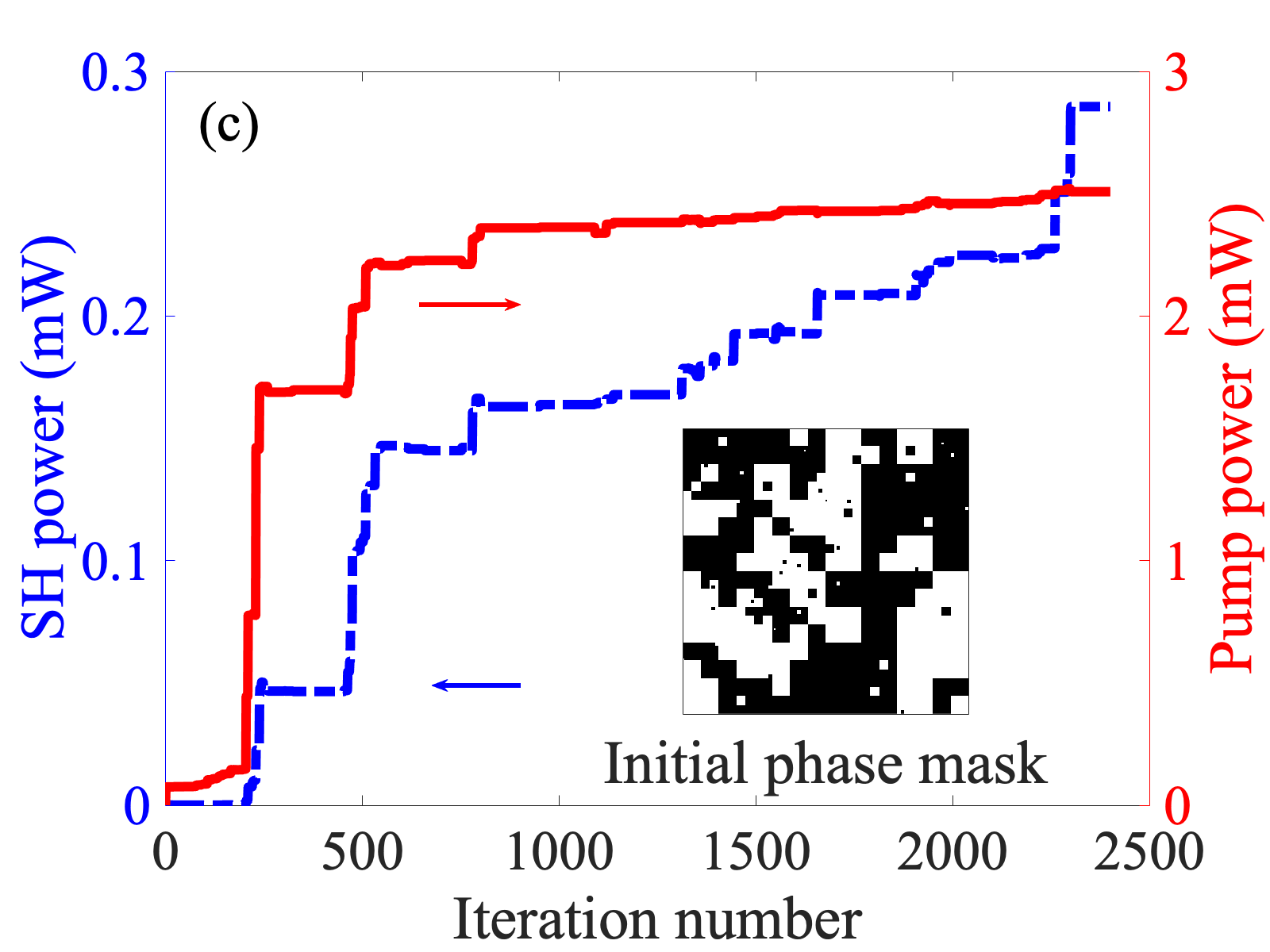}
\includegraphics[width=8.0cm]{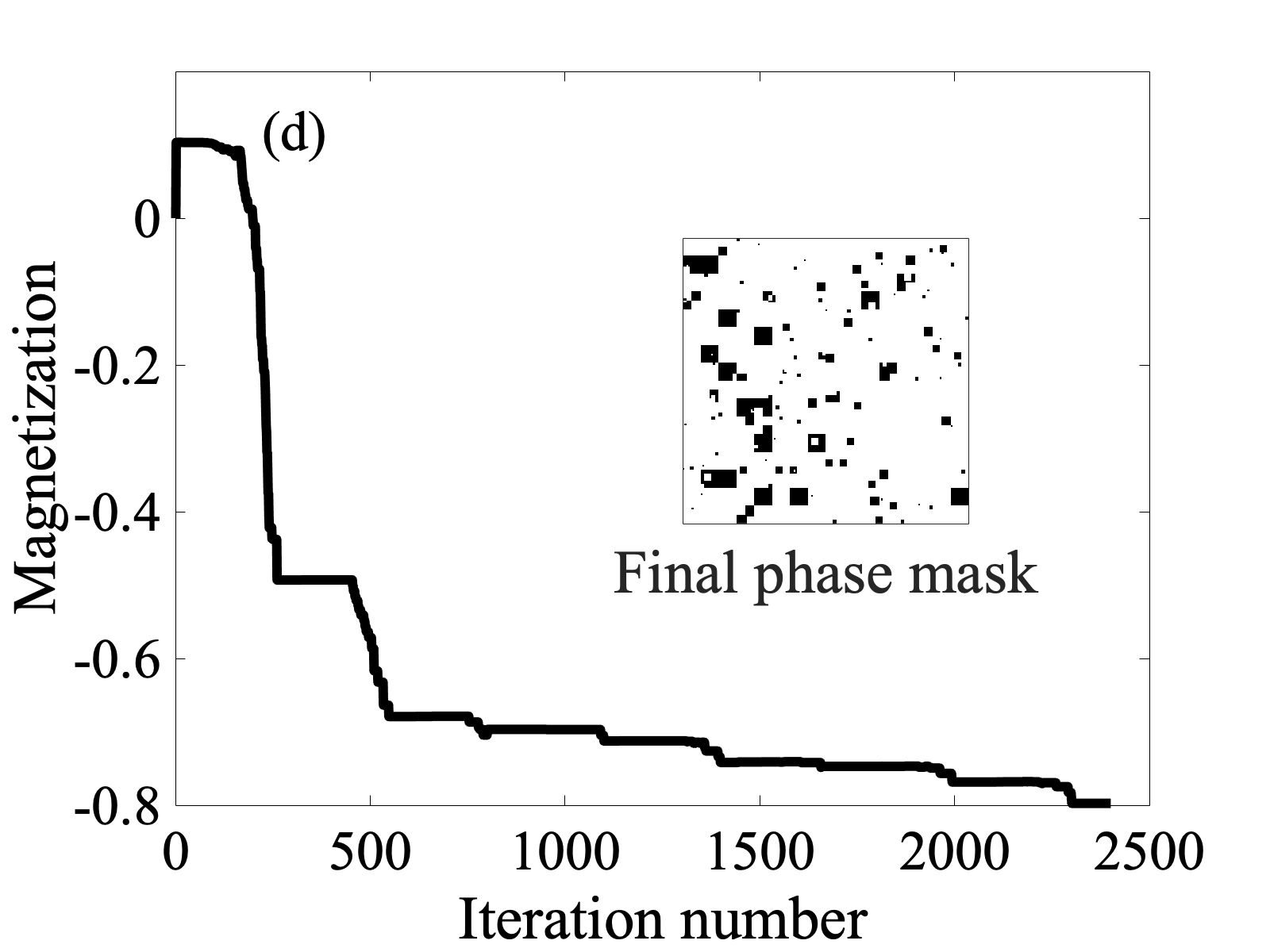}
\caption{ (a) Experimentally measured evolution of the SH (blue dashed curve) and pump (red solid curve) powers during optimization, set to find the ground state of the Ising machine with $\alpha=-1$, $\beta=-1$, and $\gamma=-1$. (b) show the evolution of the magnetization. (c) and (d) are similar results with $\alpha=1$, $\beta=-1$, and $\gamma=-1$. Inset in (a,c) and (b,d) are initial and final phase masks for 800 $\times$ 800 spins, respectively, where the black and white pixels indicates the positive and negative orientated spins, respectively. } 
\label{pic_Results_4}
\end{figure}

As an example, in Fig. \ref{pic_Results_4}, we compare the results with $\alpha=1$ and $-1$, both with $\beta=-1$ and $\gamma=-1$. The results are shown in linear scale of power vs iteration number. In both cases, the spins are start aligning as the SH and pump power increases. However, magnetization can have positive or negative orientation as can be seen in Fig. \ref{pic_Results_4}(b) and (d). Insets show the initial and final phase masks in each case, respectively, where the black and white pixels indicates the positive and negative orientated spins, respectively.

\end{document}